\documentclass[pdftex,twocolumn,epjc3]{svjour3}

\RequirePackage[T1]{fontenc}

\smartqed  

\RequirePackage{graphicx}
\RequirePackage{mathptmx}      
\RequirePackage{flushend}
\RequirePackage[numbers,sort&compress]{natbib}
\RequirePackage[colorlinks,citecolor=blue,urlcolor=blue,linkcolor=blue]{hyperref}

\journalname{Eur. Phys. J. C}

  \usepackage{graphics}
  \usepackage{float}
  \usepackage{bm}        
  \usepackage{amssymb}   
  \usepackage{slashed}   
  \usepackage{amsmath}   
  \usepackage{verbatim}  
  \usepackage{color} 
  \usepackage{xcolor} 
  \usepackage{subfig}
\begin{document}


\title{Searching for anomalous quartic gauge couplings at muon colliders using principal component analysis}








\author{Yi-Fei Dong\thanksref{e1,addr1}
        \and
        Ying-Chen Mao\thanksref{e2,addr1}
        \and
        Ji-Chong Yang\thanksref{cr,e3,addr1,addr2}
}

\institute{Department of Physics, Liaoning Normal University, Dalian 116029, China\label{addr1} 
           \and
          Center for Theoretical and Experimental High Energy Physics, Liaoning Normal University, Dalian 116029, China\label{addr2}}

\thankstext[$\star$]{cr}{Corresponding author}
\thankstext{e1}{e-mail: dyf2818051165@163.com}
\thankstext{e2}{e-mail: myc@lnnu.edu.cn}
\thankstext{e3}{e-mail: yangjichong@lnnu.edu.cn}

\date{Received: date / Revised version: date}

\maketitle

\abstract{
Searching for new physics~(NP) is one of the areas of high-energy physics that requires the most processing of large amounts of data.
At the same time, quantum computing has huge potential advantages when dealing with large amounts of data. 
The principal component analysis~(PCA) algorithm may be one of the bridges connecting these two aspects. 
On the one hand, it can be used for anomaly detection, and on the other hand, there are corresponding quantum algorithms for PCA. 
In this paper, we investigate how to use PCA to search for NP. 
Taking the example of anomalous quartic gauge couplings in the tri-photon process at muon colliders, we find that PCA can be used to search for NP. 
Compared with the traditional event selection strategy, the expected constraints on the operator coefficients obtained by PCA based event selection strategy are even better.
}




\section{\label{sec1}Introduction}

Since its establishment, the Standard Model~(SM) has withstood the challenge of high-precision experimental measurements in describing strong interaction, electromagnetic and weak interaction phenomena. 
The theoretical results are almost in perfect agreement with the experimental observations, but some important fundamental problems are still difficult to be explained within the framework of the SM~\cite{neutrinomass1,neutrinomass2,g2muon,rdstar2,wmass}. 
The seeking for new physics~(NP) beyond the SM has become one of the most advanced and important topics in high energy physics~(HEP)~\cite{johnellis}.

As the colliders' performance improves and their luminosities increase, the efficient ways to process large amounts of data become important.
One of the efficient ways is to use machine learning~(ML) algorithms.
ML is a general term for a class of algorithms, including probability theory, statistics and other disciplines. 
Because of its advantages in complex data processing, ML algorithms have been applied in many fields, including HEP~\cite{mlreview,ml1,ml2,ml3,ml4,ml5,ml6,ml8,ml10,ml11,wwww,wwwwunitary}. 
At present, many applications show that the anomaly detection~(AD) ML method can be effectively applied in the phenomenological studies of NP. 
One of the advantages is that, when AD is applied to search for NP, the implementation is often independent of the NP model to be searched~\cite{ad,guassian,autoencoder1,autoencoder2,ml7,ml9,ml12,Zhang:2023yfg}.
Nevertheless, it should be pointed out that the tunable parameters related to AD methods are often dependent on the NP models and processes to be studied.

In addition, quantum computing is another efficient way to process large amounts of complex data.
Many ML algorithms can be accelerated by quantum computing~\cite{qml1,qml2,Garcia:2022cqq}.
One example is the principal component analysis~(PCA) algorithm~\cite{pca,qpca}. 
PCA algorithm can also be used in AD, but it is not clear whether PCA anomaly detection~(PCAAD) algorithm is useful for searching NP.
If it was the case, then it is strongly implied that quantum PCA can also be used to search for NP signals and we will probably get a way to discover NP using quantum computers.
There are other examples such as the autoencoder~(AE)~\cite{LIOU20083150,LIOU201484,vae1,vae2}.
AE is better than PCA in handling data dimensionality reduction, so it has a good potential to perform better than PCA in AD as well~\cite{autoencoder1,autoencoder2}. 
Not only that, AE also has the potential for quantum acceleration~\cite{Romero_2017,Bravo-Prieto_2021,PhysRevLett.124.130502,qvae}. 
However, PCA is more explicit in a geometric sense, as the events are mainly distributed along the eigenvectors of the covariance matrix.
Moreover, PCAAD does not require the cooperation of other algorithms.
Therefore, we focus on the PCAAD in this paper and leave the study of the AE for the future.

To test whether the PCAAD algorithm is feasible, we intend to conduct experiments by searching for dimension-8 operators in the SM effective field theory~(SMEFT) contributing to anomalous quartic gauge couplings~(aQGCs).
Note that there are already ML approaches targeting SMEFT which have been shown to able to enhance the signal significance~\cite{Chen:2020mev,ml6,ml7,wwww,wwwwunitary}.

There are many reasons to consider dimension-8 operators contributing to aQGCs.
For example, the dimension-8 operators are important w.r.t. the convex geometry perspective to the operator space~\cite{positivity1,positivity2,positivity3}.
Moreover, there exist various NP models generating dimension-8 effective operators relevant for aQGCs~\cite{composite1,composite2,extradim,2hdm1,2hdm2,zprime1,zprime2,alp1,alp2,wprime}, and there are distinct cases where dimension-6 operators are absent but the dimension-8 operators show up~\cite{bi1,bi2,bi3}.
As a result, aQGCs have received a lot of attention~\cite{d81,vbs1,ssww,wastudy,wwstudy,zastudy,sswwexp1,sswwexp2,zaexp1,zaexp2,zaexp3,waexp1,zzexp1,zzexp2,wzexp1,wzexp2,wwexp1,wwexp2,wvzvexp,waexp2,zzexp3}.
The existence of aQGCs makes the tri-photon process at the muon colliders inconsistent with the predictions of the SM~\cite{triphoton}. 
Recently, the studies of muon colliders also have drawn a lot of attention~\cite{muoncollider1,muoncollider2,muoncollider4,muoncollider6,muoncollider7,muoncollider8,muoncollider3,muoncollider5,Han:2020uid,Han:2021kes,Aime:2022flm,Yin:2020afe}.
Although, the PCAAD might be suitable for various cases, as a test bed, we use the PCAAD to search for the signals of aQGCs in the tri-photon process at the muon colliders.

The rest of this paper consists of the following.
In Sec.~\ref{sec2}, the aQGCs and the tri-photon process are briefly reviewed. 
Event selection strategy using PCAAD is discussed in Sec.~\ref{sec3} with the contribution of interference ignored and the contributions of the SM and NP considered separately because we focus on the features of the SM and NP events in this section.
In our approach the eigenvectors are solely determined by the SM events.
Sec.~\ref{sec4} presents the expected constraints on operator coefficients.
In Sec.~\ref{sec4}, the interference terms between the SM and the NP are taken into account.
Sec.~\ref{sec5} summarizes our main conclusions.

\section{\label{sec2}The effect of aQGCs in the tri-photon process}

\begin{table}[hbtp]
\centering
\begin{tabular}{c|c}
\hline
coefficient & constraint \\
\hline
$f_{T_{0}}/\Lambda^4$ &[-0.12, 0.11]~\cite{wvzvexp} \\
$f_{T_{1}}/\Lambda^4$ &[-0.12, 0.13]~\cite{wvzvexp} \\
$f_{T_{2}}/\Lambda^4$ &[-0.28, 0.28]~\cite{wvzvexp} \\
$f_{T_{5}}/\Lambda^4$ &[-0.5, 0.5]~\cite{waexp2}\\
$f_{T_{6}}/\Lambda^4$&[-0.4, 0.4]~\cite{waexp2} \\
$f_{T_{7}}/\Lambda^4$&[-0.9, 0.9]~\cite{waexp2} \\
$f_{T_{8}}/\Lambda^4$&[-0.43, 0.43]~\cite{zzexp3} \\
$f_{T_{9}}/\Lambda^4$&[-0.92, 0.92]~\cite{zzexp3}\\
\hline
\end{tabular}
\caption{Constraints on $O_{T_{i}}$ coefficients at $95~\%$ C.L obtained by the LHC.}
\label{table:1}
\end{table}

Because only the transverse operators $O_{T_i}$ contribute to the tri-photon process~\cite{triphoton}, in this paper, we consider only $O_{T_i}$, they are~\cite{aqgcold,aqgcnew}
\begin{equation}
\begin{split}
&O_{T,0}={\rm Tr}\left[\widehat{W}_{\mu\nu}\widehat{W}^{\mu\nu}\right]\times {\rm Tr}\left[\widehat{W}_{\alpha\beta}\widehat{W}^{\alpha\beta}\right],\\
&O_{T,1}={\rm Tr}\left[\widehat{W}_{\alpha\nu}\widehat{W}^{\mu\beta}\right]\times {\rm Tr}\left[\widehat{W}_{\mu\beta}\widehat{W}^{\alpha\nu}\right],\\
&O_{T,2}={\rm Tr}\left[\widehat{W}_{\alpha\mu}\widehat{W}^{\mu\beta}\right]\times {\rm Tr}\left[\widehat{W}_{\beta\nu}\widehat{W}^{\nu\alpha}\right],\\
&O_{T,5}={\rm Tr}\left[\widehat{W}_{\mu\nu}\widehat{W}^{\mu\nu}\right]\times B_{\alpha\beta}B^{\alpha\beta},\\
&O_{T,6}={\rm Tr}\left[\widehat{W}_{\alpha\nu}\widehat{W}^{\mu\beta}\right]\times B_{\mu\beta}B^{\alpha\nu},\\
&O_{T,7}={\rm Tr}\left[\widehat{W}_{\alpha\mu}\widehat{W}^{\mu\beta}\right]\times B_{\beta\nu}B^{\nu\alpha},\\
&O_{T,8}=B_{\mu\nu}B^{\mu\nu}\times B_{\alpha\beta}B^{\alpha\beta},\\
&O_{T,9}=B_{\alpha\mu}B^{\mu\beta}\times B_{\beta\nu}B^{\nu\alpha},\\
\end{split}
\label{eq.2.1}
\end{equation}
where $\widehat{W}\equiv \vec{\sigma}\cdot {\vec W}/2$ , $\sigma$ is the Pauli matrices , ${\vec W}=\{W^1,W^2,W^3\}$, $B_{\mu}$ and $W_{\mu}^i$ are $U(1)_{\rm Y}$ and $SU(2)_{\rm I}$ gauge fields, $B_{\mu\nu}$ and $W_{\mu\nu}$ correspond to the field strength tensors. Table~\ref{table:1} shows the constraints on the coefficients obtained by the large hadron collider~(LHC).

\begin{figure}[htbp]
\resizebox{1.0\hsize}{!}{\includegraphics*{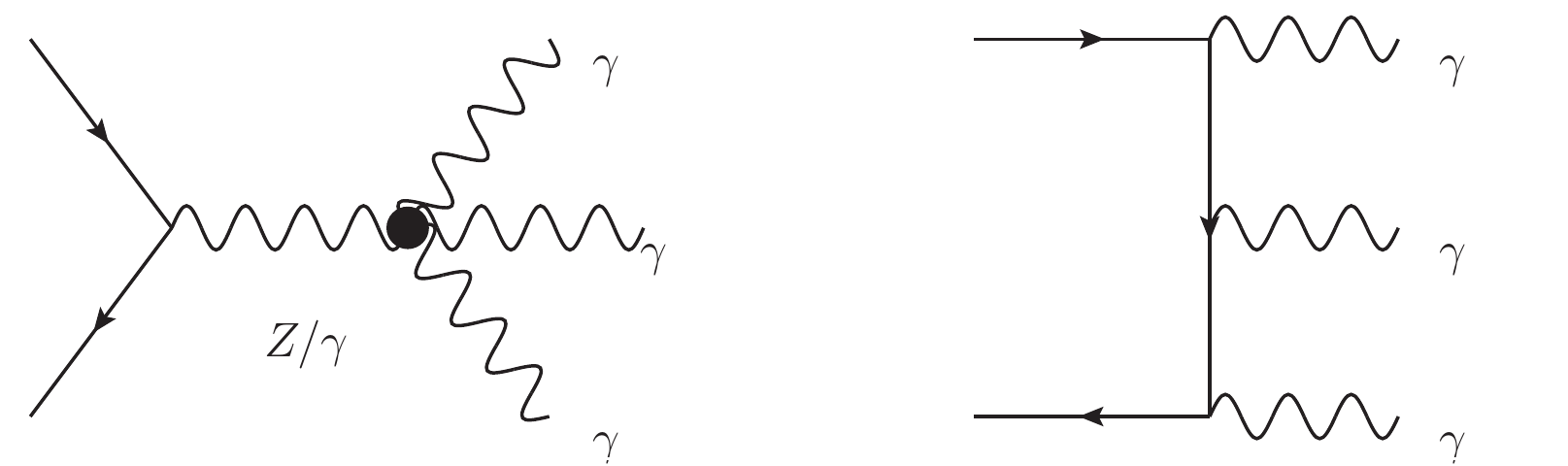}}
\caption{\label{fig:Feyman}Feynman diagrams for the process $\mu^{+} \mu^{-} \rightarrow \gamma\gamma\gamma$. The process derived by the $O_{T_{i}}$ operator is on the left, the representative diagram of SM is on the right. In the case of the SM, permuting the photons in final state can arrive in other five diagrams.}
\end{figure}

At muon colliders, which are also called gauge boson colliders~\cite{muoncollider5}, it is expected that, the vector boson scattering/fusion~(VBS/VBF) processes are dominant production modes for both the SM and NP starting from a few TeV energies because of the logarithmic enhancement from gauge boson radiation~\cite{muoncollider3,muoncollider5,Han:2020uid}.
For $O_{T_i}$ operators, the dimension of cross section indicates $\sigma _{NP}\sim s^3 (f/\Lambda^4)^2$, except for logarithmic enhancement, the contribution for both VBS and tri-boson grow by $s^3$, i.e., the momentum dependence in the Feynman rule of the aQGCs cancels the $1/s$ in the propagator in the tri-boson process.
Besides, different from the LHC, where the tri-boson contributions from the aQGCs are further suppressed because they must be led by sea quark partons, no such suppression occurs at muon colliders except for that the s-channel propagator could not be the $W$ bosons.
It has been shown that, at muon colliders, for $O_{T_5}$ the tri-boson contribution is competitive to the VBS for the process $\mu^+\mu^-\to \gamma\gamma \nu\bar{\nu}$, it surpasses the VBS when $\sqrt{s}<5\;{\rm TeV}$, and is about $1/3$ of the VBS when $\sqrt{s}=30\;{\rm TeV}$~\cite{triphoton}.
Moreover, for the tri-photon process, there are no subsequent decays in final states.
As a result, the sensitivity of the tri-photon process at the muon colliders to the $O_{T_i}$ operators is competitive to or even better than the VBS processes.
For simplicity, we consider only the tri-photon process in this paper.

The left panel of Fig.~\ref{fig:Feyman} shows the Feynman diagrams induced by $O_{T_i}$ operators, while the right panel shows the SM background.
The contributions of $O_{T_{1,6}}$ operators are exactly equal to $O_{T_{0,5}}$ operators, respectively. 
Therefore, in the following we consider only $O_{T_{0,2,5,7,8,9}}$ operators.

\section{\label{sec3}PCA assistant event selection strategy}

PCA algorithm is one of the most commonly used linear dimensionality reduction methods.
It uses mathematical dimensionality reduction or feature extraction to linearly transform a large number of original variables to a smaller number of important variables, which can explain the main information of the original data-set. 

Denoting the points in a data-set as $\vec{p}_i=(p_i^1, p_i^2, \ldots, p_i^m)$, where $1 \leq i \leq N$, $N$ is the number of points in the data-set and $m$ is the number of dimensions of the points~(which is also called the number of features of the points), the PCA algorithm start with a standardization of the points.
In this paper, we use z-score standardization, $x_i^j=(p_i^j - \bar{p}^j)/\epsilon ^j$, where $\bar{p}^j$ and $\epsilon ^j$ are the mean and standard deviation of the $j$-th feature for the whole data-set.
Then the data-set can be expressed as a matrix $X=(\vec{x}_1^T, \vec{x}_2^T, \ldots, \vec{x}_N^T)$.
An $m\times m$ covariance matrix $C$ can be obtained as $C=XX^T$.
Next, we need to calculate the eigenvectors of $C$.
Denoting $\vec{\eta}_j$ as the eigenvector corresponding to eigenvalue $\lambda _j$, where $\{\lambda _j\}$ are in descending order.
The PCA algorithm project the points in the data-set using $\vec{\eta}_j$ to new features, hence the new features are $\tilde{x}_i^j = \vec{\eta}_j \cdot \vec{x}_i$.

The process of reducing the original data dimension by PCA can be equivalent to projecting the original features onto the dimension with the maximum amount of projection information.
Each eigenvector $\vec{\eta}_j$ is a projection axis, and each $\vec{\eta}_j$ corresponding eigenvalue $\lambda_j$ is the variance of the original features projected onto the projection axis. 
According to the maximum variance theory, the larger the variance, the larger the information, after projection, in order to ensure that the information lost is small, one have to choose the projection axis with larger variance, that is, to choose the $\vec{\eta}_j$ corresponding to the larger $\lambda_j$.
By selecting $m'$ eigenvectors corresponding to the first $m'$ eigenvalues, the dimension of the points is reduced from $m$ for $\vec{x}_i$ to $m'$ for $\vec{\tilde{x}}_i$.
The components of $\vec{\tilde{x}}_i$, i.e. $\tilde{x}_i^j$ with $1\leq j \leq m'$ are the first $m'$ principal components of the data-set.

The eigenvector reflects the directions corresponding to the degree of variance change of the original data, and the eigenvalue is the variance of the data in the corresponding direction. 
Since the background is mainly distributed along the most important eigenvectors, it can be expected that points in the background are closer to the eigenvectors after decentralization.
And if the signal does not coincide with the background distribution, then the distances between the signal points and eigenvectors will be larger.
Therefore, after the $\vec{\eta}_j$ are obtained, we use the distance between the points and $\vec{\eta}_j$~(denoted as $d_{i,j}$) as anomaly scores to search for the signal events.

\subsection{\label{sec3.1}Data preparation}

To test the feasibility of the PCAAD algorithm for searching for aQGCs, we build the data-set using Monte Carlo~(MC) simulation with the help of \verb"MadGraph5@NLO" toolkit~\cite{madgraph,feynrules,ufo}, and a muon collider-like detector simulation with \verb"Delphes"~\cite{delphes} is applied. 
The data preparation is applied by \verb"MLAnalysis"~\cite{Guo:2023nfu}. 
To avoid infrared divergences, in this section we use the standard cuts as default, and the cuts associated with infrared divergences are
\begin{equation}
\begin{split}
&p_{T,\gamma} > 10\;{\rm GeV},\; |\eta _{\gamma}| < 2.5, \; \Delta R_{\gamma\gamma} > 0.4,
\end{split}
\label{eq.standardcuts}
\end{equation}
where $p_{T,\gamma}$ and $\eta _{\gamma}$ are the transverse momentum and pseudo-rapidity for each photon, respectively, $\Delta R_{\gamma\gamma}=\sqrt{\Delta \phi ^2 + \Delta \eta ^2}$, where $\Delta \phi$ and $\Delta \eta$ are the differences in azimuth angles and pseudo-rapidities between two photons, respectively. 
The events of the signal are generated one operator at a time. 
In this section, we choose the coefficients as upper bounds listed in Table~\ref{table:1}.
We require that each event consists of at least three photons, so that an event in the data-set consists of $12$ numbers which are the components of the four momenta of three photons, which means that an event corresponds to a dimension-12 vector.
In this paper, the hardest three photons are selected, and the photons are arranged in descending order of energy for each event.
In this section, we generate 600000 events for the SM and 30000 events for NP, respectively. 
It needs to point out that, the interference is ignored in this section because we concentrate on the features of the background and signal.

Note that we do not use the physical information in the data. 
This is because we want to verify that our method is independent of the physical content.
For example, for photons, the four components of the four-momentum are not independent of each other, and the four-momenta of three photons are not independent of each other. 
We assume that the PCAAD does not know that these data represent the four-momenta of the photons, and do not use the above relationships, but only treat the data as vectors with 12 numbers.

\subsection{\label{sec3.2}Event selection strategy}

To search for NP in the target data-set, based on PCAAD, we draw out the following procedure,
\begin{enumerate}
\item Prepare a data-set for the SM using MC. The data-set of the SM is denoted as $p^{\rm SM}$, and the data-set of the target is denoted as $p^{\rm tar}$.
\item Apply the z-score standardization to the data-sets of the SM $x_i^{\rm SM,j}=(p_i^{\rm SM,j} - \bar{p}^{\rm SM,j})/\epsilon ^{\rm SM,j}$, where $\bar{p}^{\rm SM,j}$ and $\epsilon ^{\rm SM,j}$ are the mean and the standard deviation of the $j$-th feature for the SM data-set. However, when standardizing the target data-set, we also use $\bar{p}^{{\rm SM},j}$ and $\epsilon ^{{\rm SM},j}$, so that $x_i^{{\rm tar},j}=(p_i^{{\rm tar},j} - \bar{p}^{{\rm SM},j})/\epsilon ^{{\rm SM},j}$.
\item Denote $X^{\rm SM}=((\vec{x}_1^{\rm SM})^T, (\vec{x}_2^{\rm SM})^T, \ldots, (\vec{x}_{N^{\rm SM}}^{\rm SM})^T)$, and $X^{\rm tar}=((\vec{x}_1^{\rm tar})^T, (\vec{x}_2^{\rm tar})^T, \ldots, (\vec{x}_{N^{\rm tar}}^{\rm tar})^T)$, where $N^{\rm SM,tar}$ are the total numbers of points in the SM and target data-sets, respectively. Calculate the covariance matrix of the SM as $C^{\rm SM}=X^{\rm SM}(X^{\rm SM})^T$.
\item Find out the eigenvalues and eigenvectors of the covariance matrix $C^{\rm SM}$.
\item Sort the eigenvalues in descending order and select the eigenvectors $\vec{\eta} _j^{\rm SM}$ corresponding to the top $m'$ ($m'< m=12$) largest eigenvalues. As will be explained later, in this paper, we use $m'=4$.
\item Find out the $m'$ new features by projecting the data-sets of both $X^{\rm SM}$ and $X^{\rm tar}$ with $\vec{\eta} _j^{\rm SM}$, i.e. $\tilde{x}_i^{\rm SM, j}=\vec{\eta} _j^{\rm SM}\cdot \vec{x}_i^{\rm SM}$, $\tilde{x}_i^{\rm tar, j}=\vec{\eta} _j^{\rm SM} \cdot \vec{x}_i^{\rm tar}$.
\item The distance from a point $\vec{x}_i$ to an eigenvector can be obtained as $d_{i,j} = \sqrt{|\vec{x}_i|^2 -\left(\vec{x}_i\cdot \vec{\eta}_j\right)^2 }= \sqrt{|\vec{x}_i|^2 -(\tilde{x}_i^j)^2 }$. $d_{i,j}$ for the points in the SM data-set and target data-set are denoted as $d_{i,j}^{\rm SM}$ and $d_{i,j}^{\rm tar}$, respectively.
\item Set cuts on $d_{i,j}$ to select events.
\end{enumerate}

It can be seen that, the PCAAD event selection strategy requires a data-set of the SM, therefore is a supervised machine learning algorithm.
In this section, we use the data-set of NP as the target data-set.

\begin{figure}[htbp]
\begin{center}
\resizebox{1.0\hsize}{!}{\includegraphics{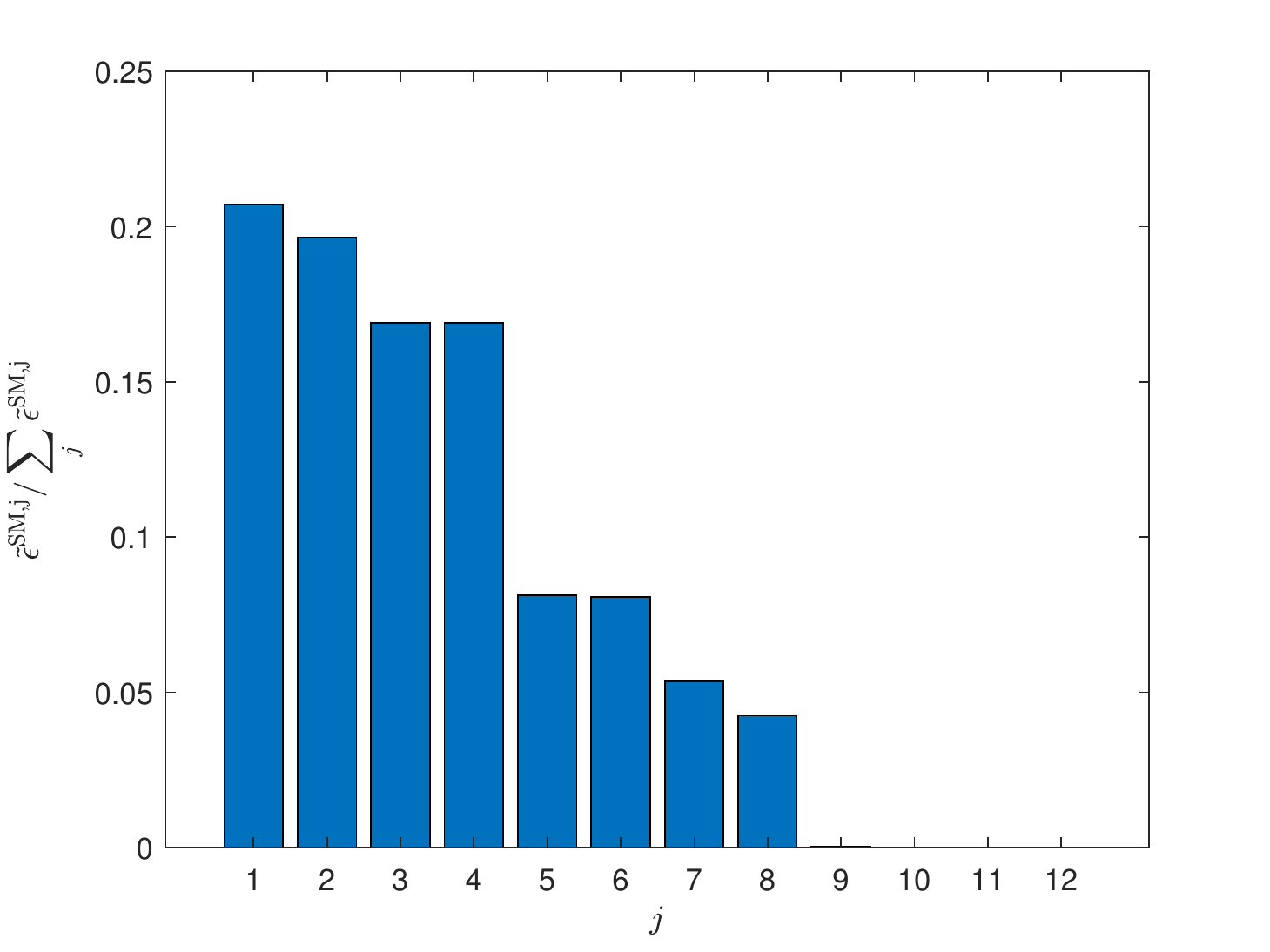}}
\caption{\label{fig:Variances}The variances of the principal components}
\end{center}
\end{figure}

\begin{figure*}[htbp]
\begin{center}
\resizebox{0.24\textwidth}{!}{\includegraphics{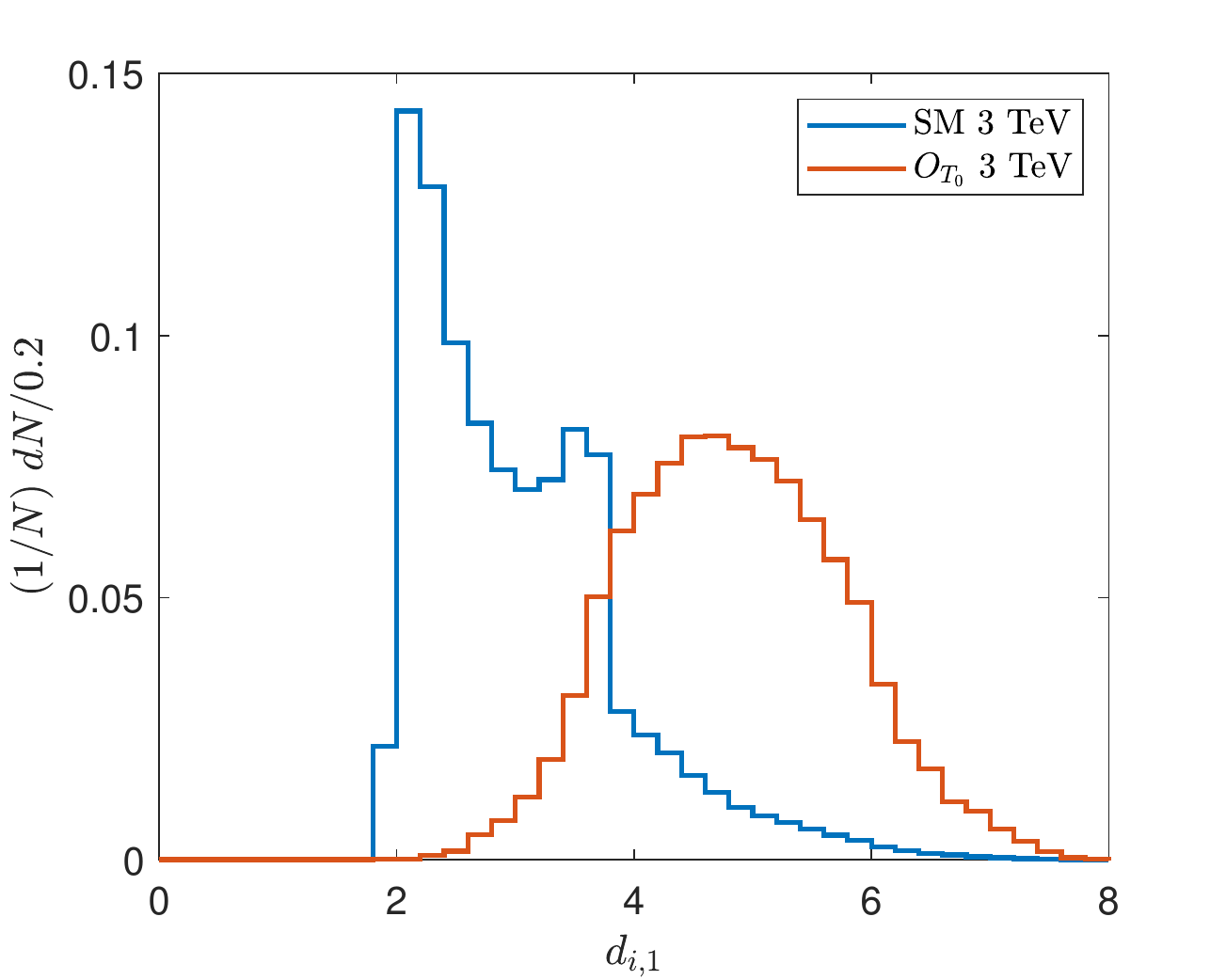}}
\resizebox{0.24\textwidth}{!}{\includegraphics{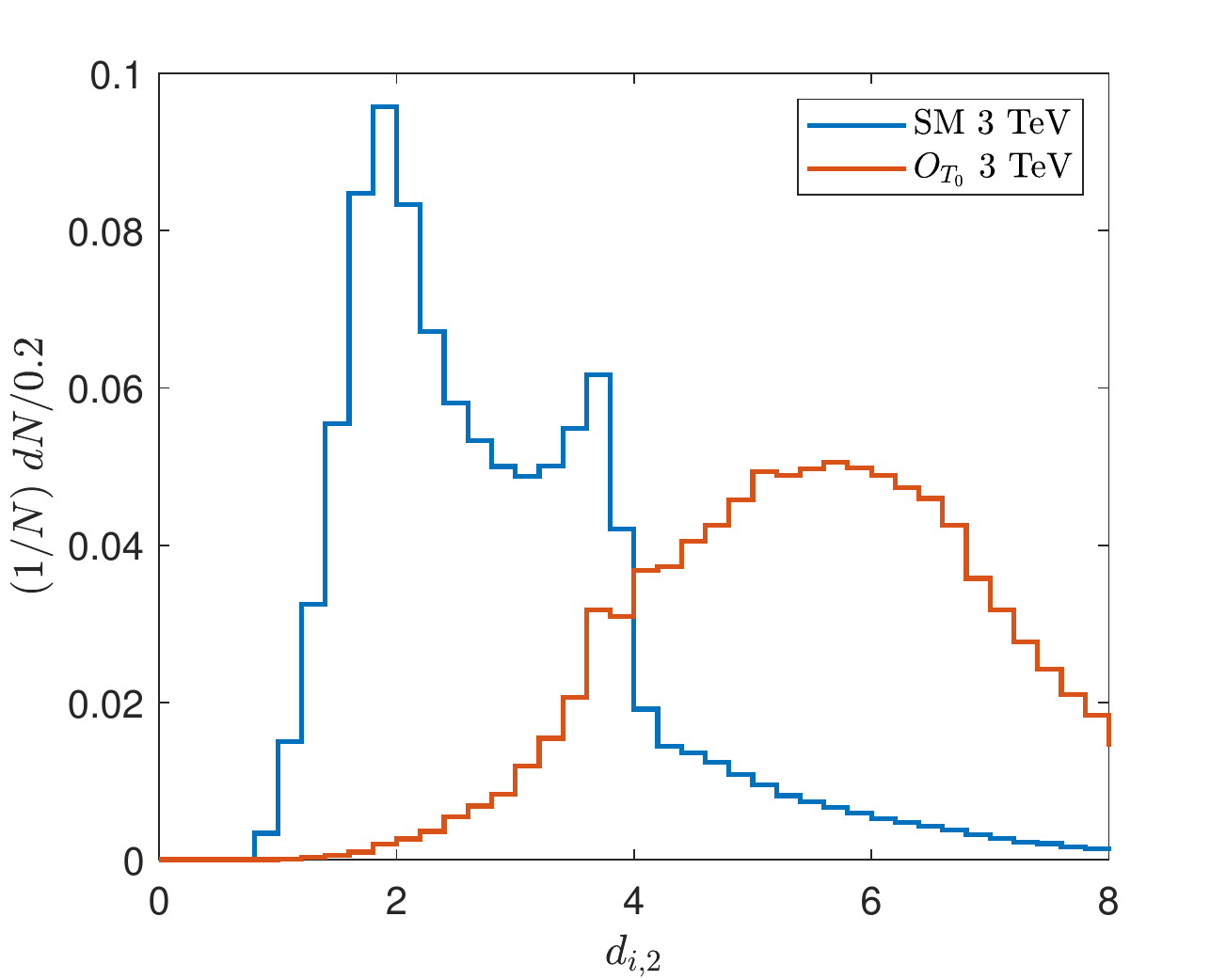}}
\resizebox{0.24\textwidth}{!}{\includegraphics{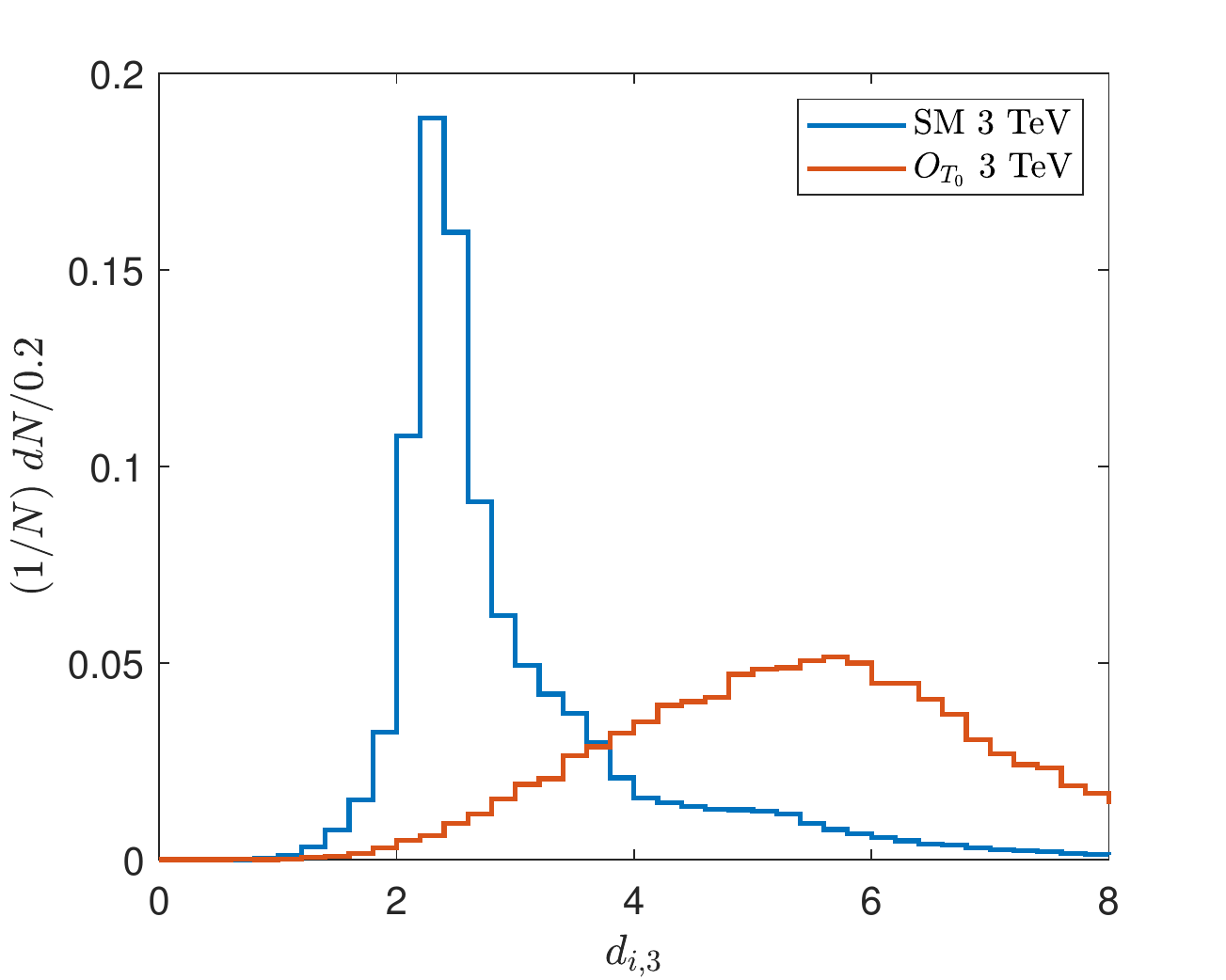}}
\resizebox{0.24\textwidth}{!}{\includegraphics{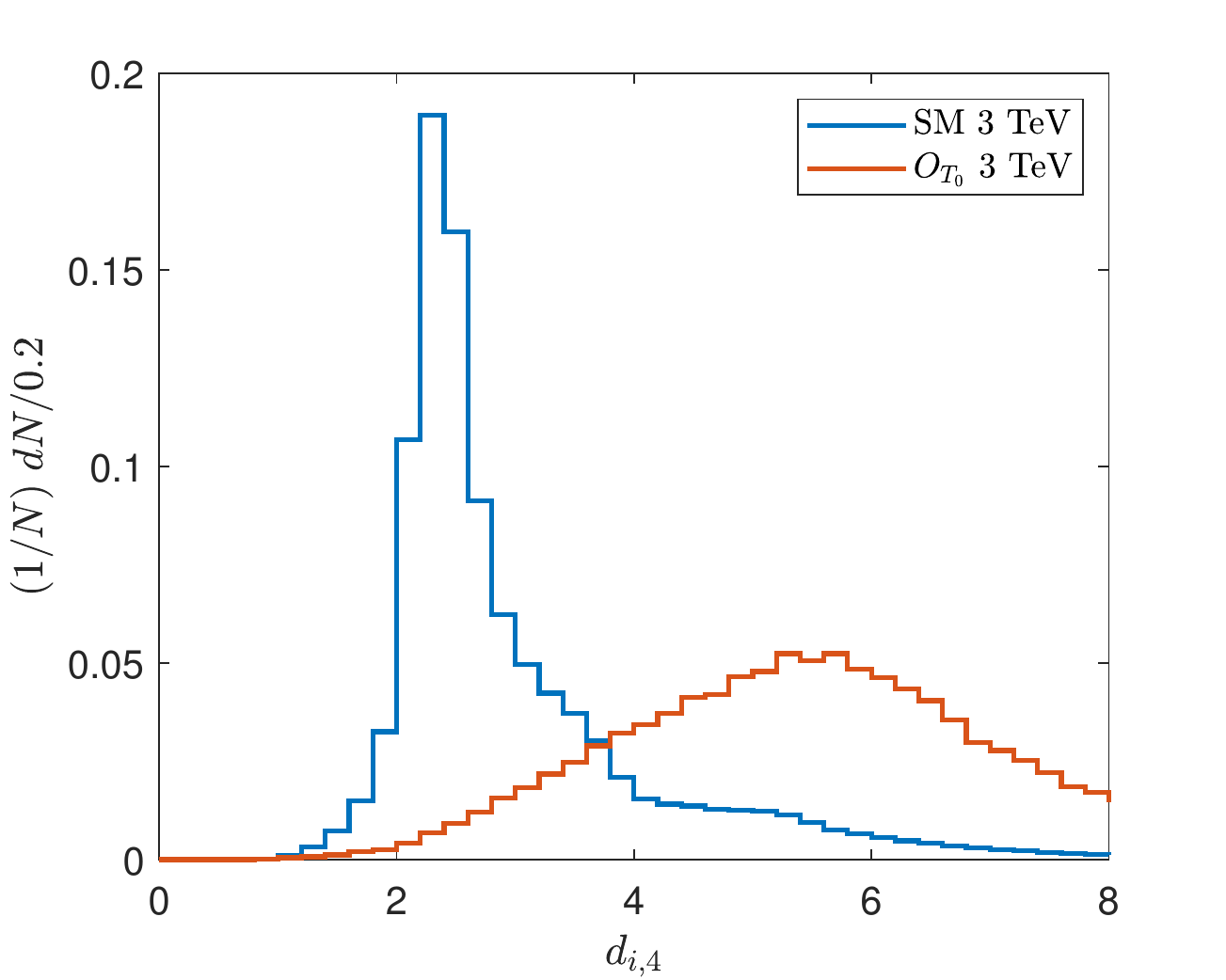}}
\resizebox{0.24\textwidth}{!}{\includegraphics{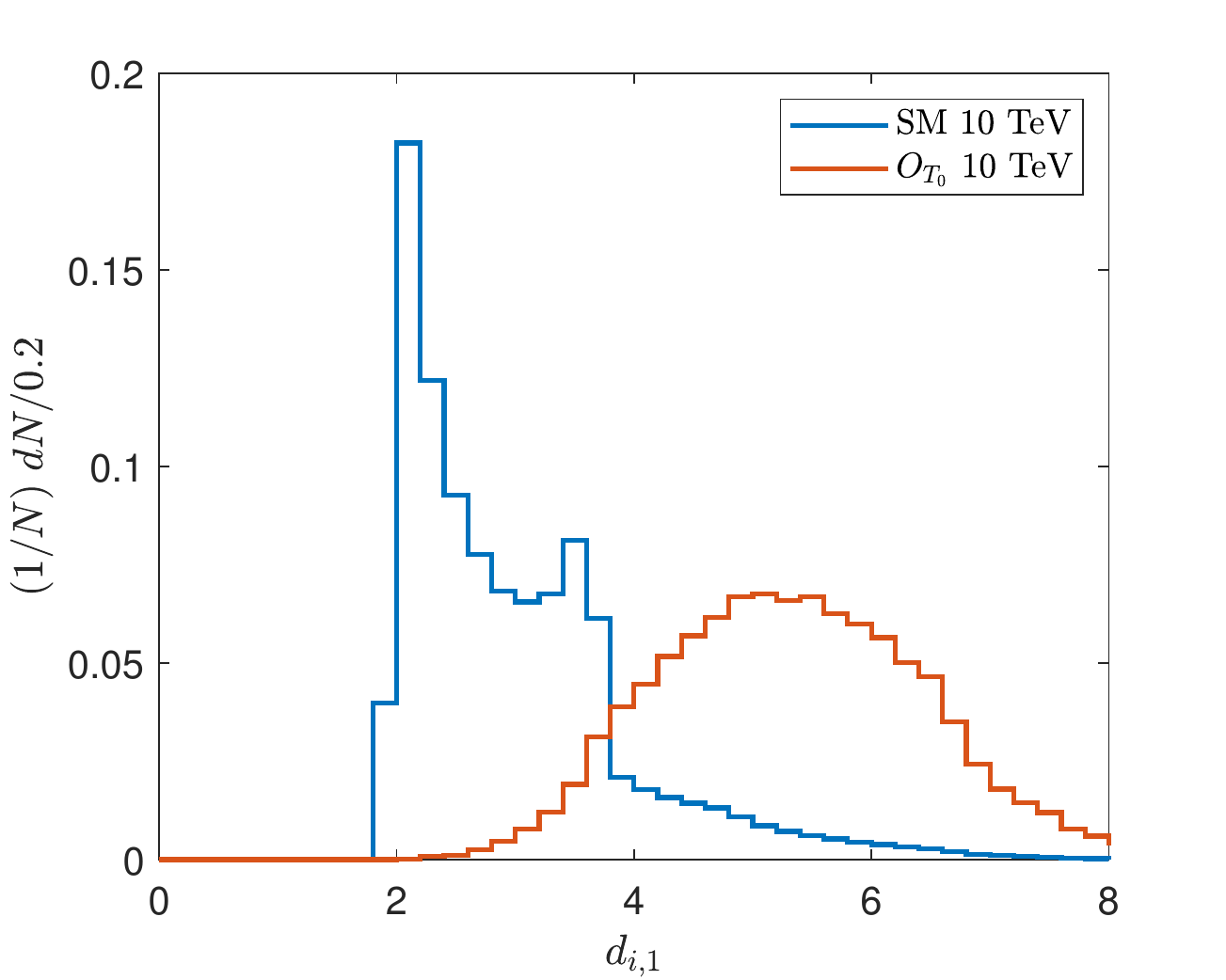}}
\resizebox{0.24\textwidth}{!}{\includegraphics{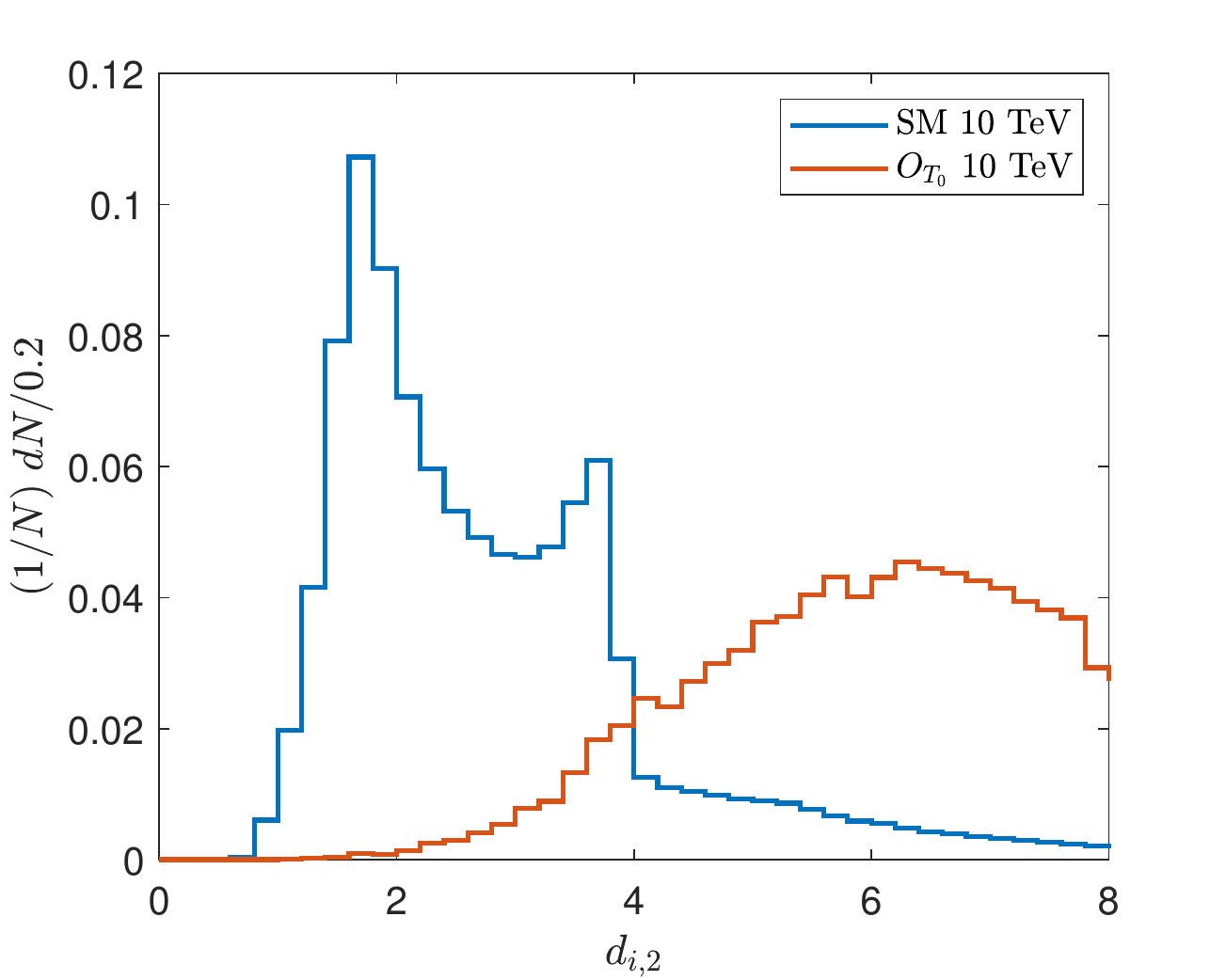}}
\resizebox{0.24\textwidth}{!}{\includegraphics{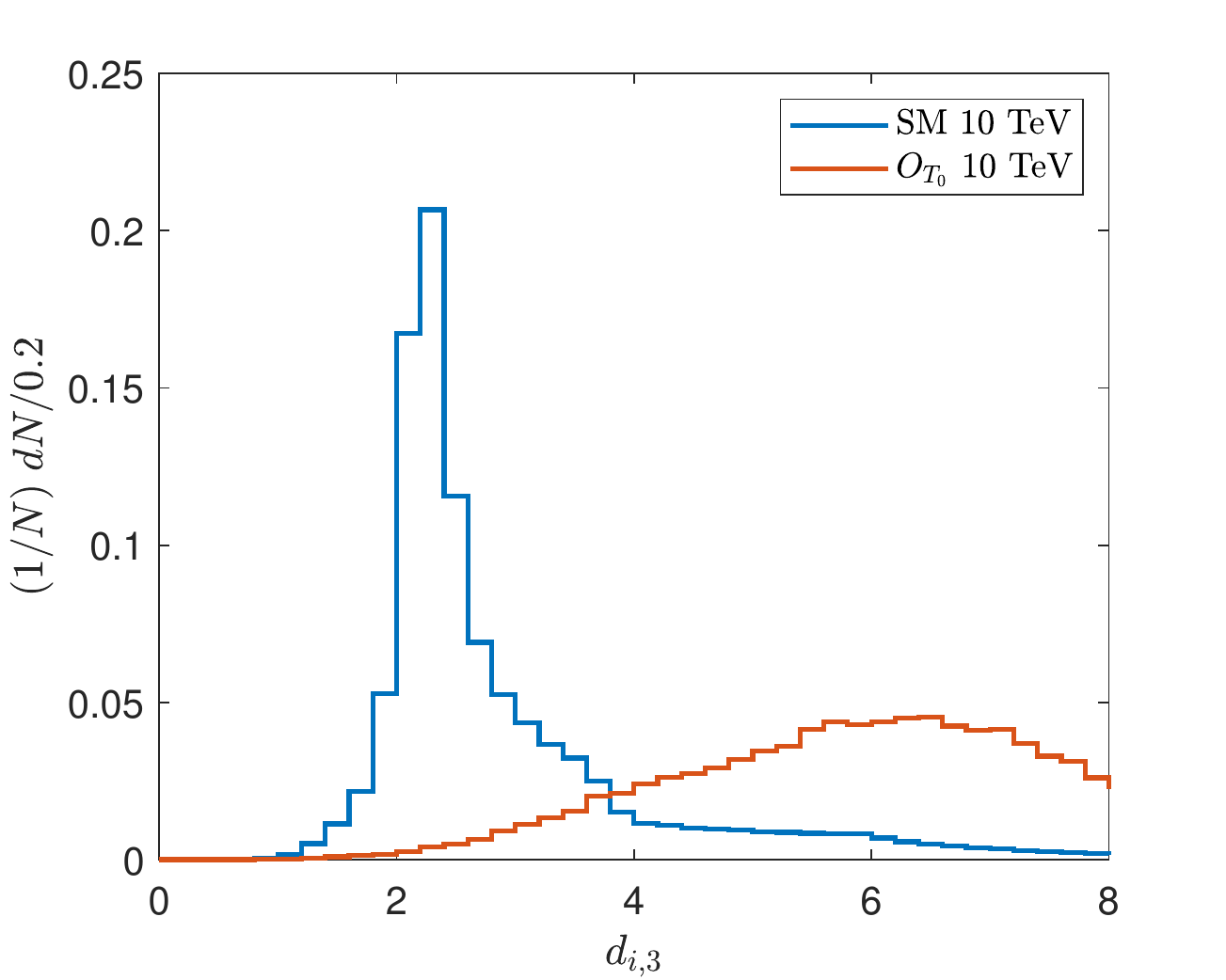}}
\resizebox{0.24\textwidth}{!}{\includegraphics{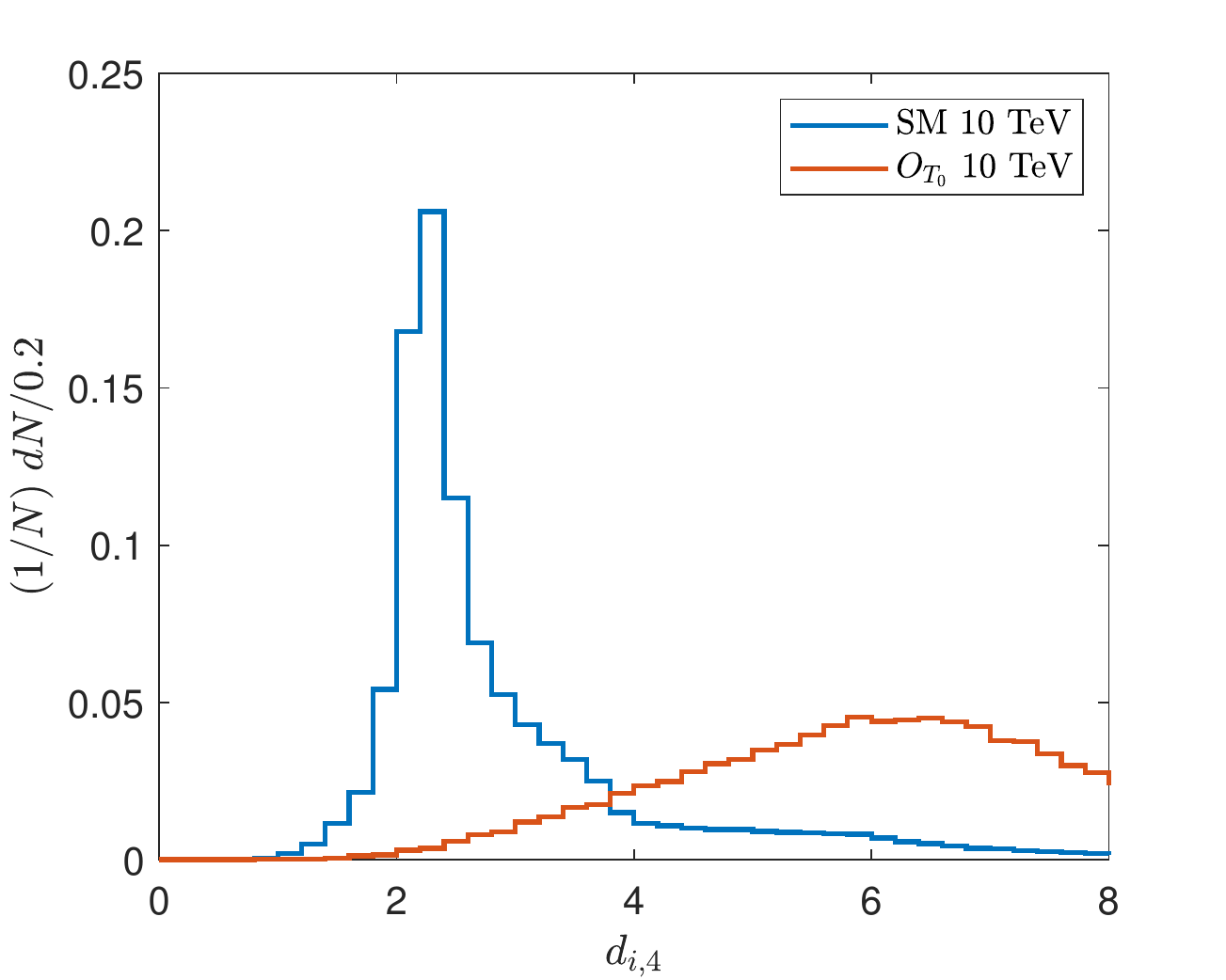}}
\resizebox{0.24\textwidth}{!}{\includegraphics{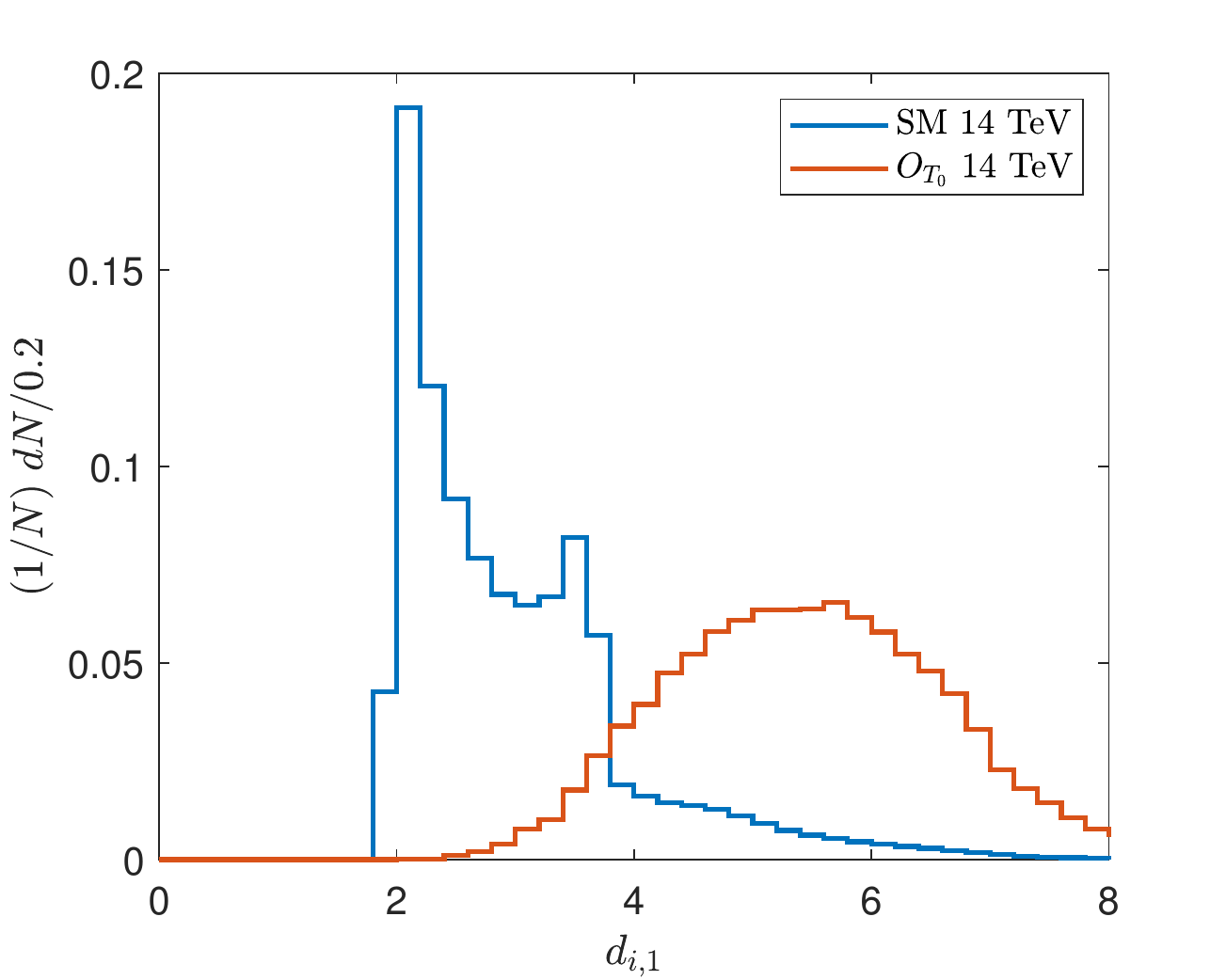}}
\resizebox{0.24\textwidth}{!}{\includegraphics{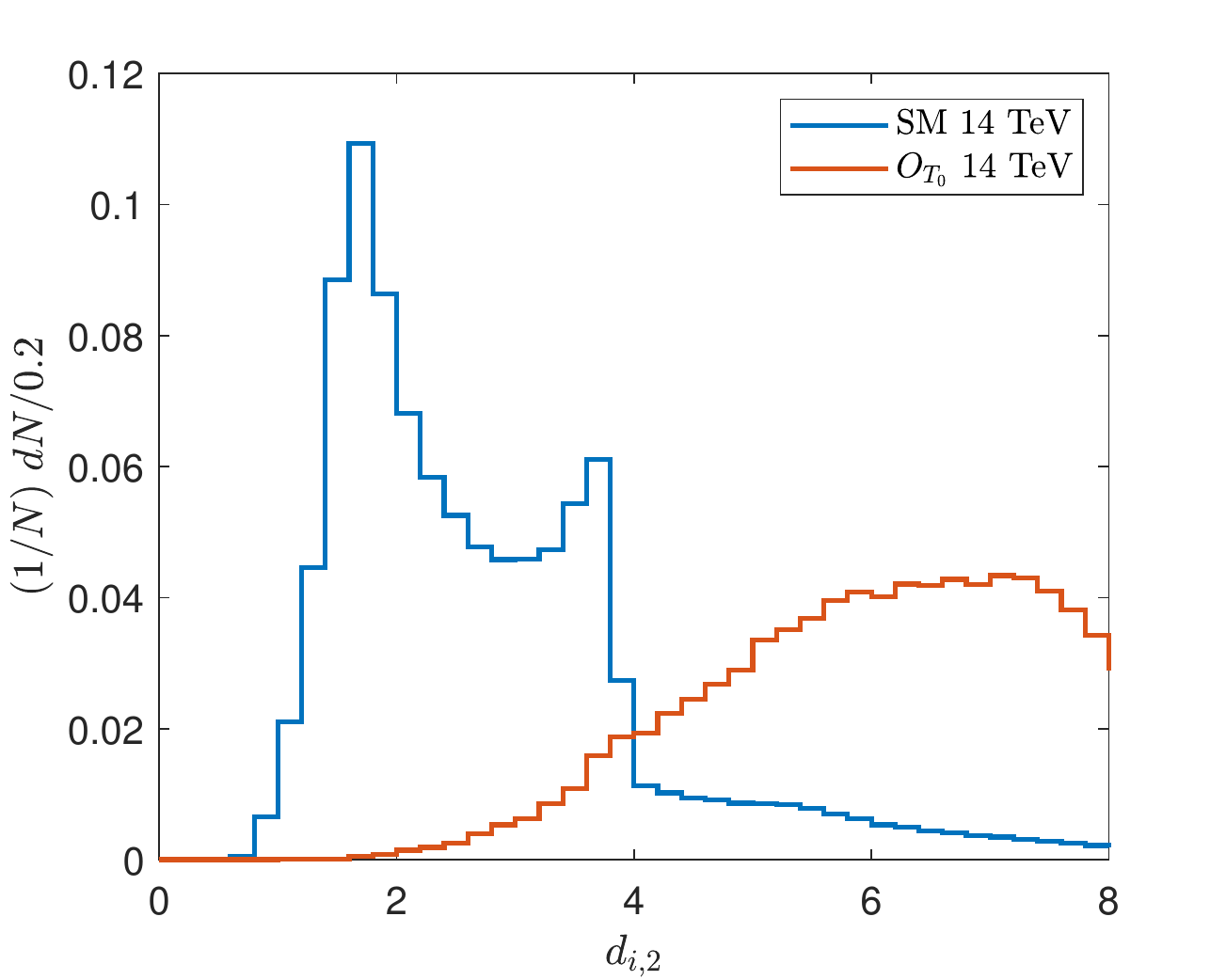}}
\resizebox{0.24\textwidth}{!}{\includegraphics{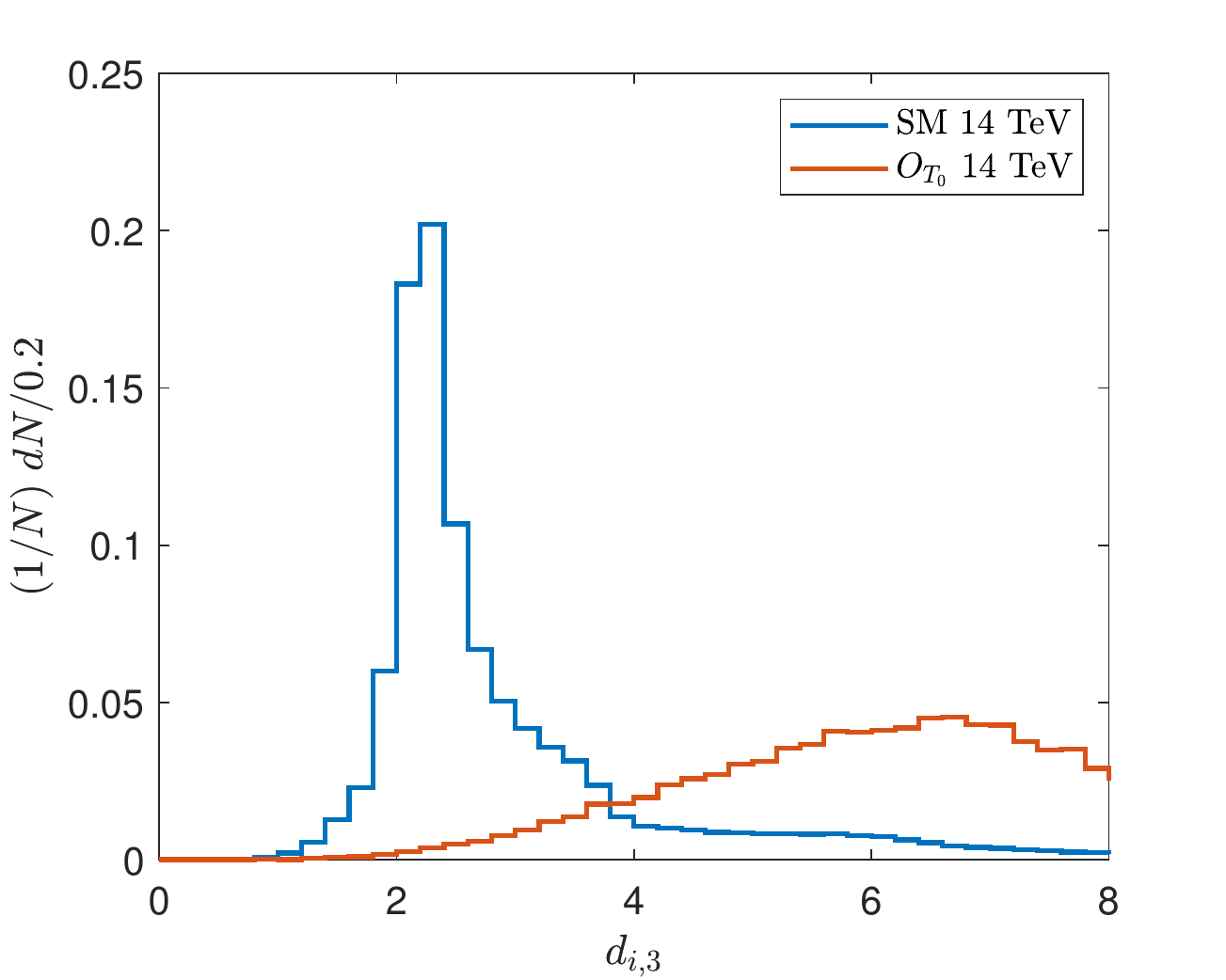}}
\resizebox{0.24\textwidth}{!}{\includegraphics{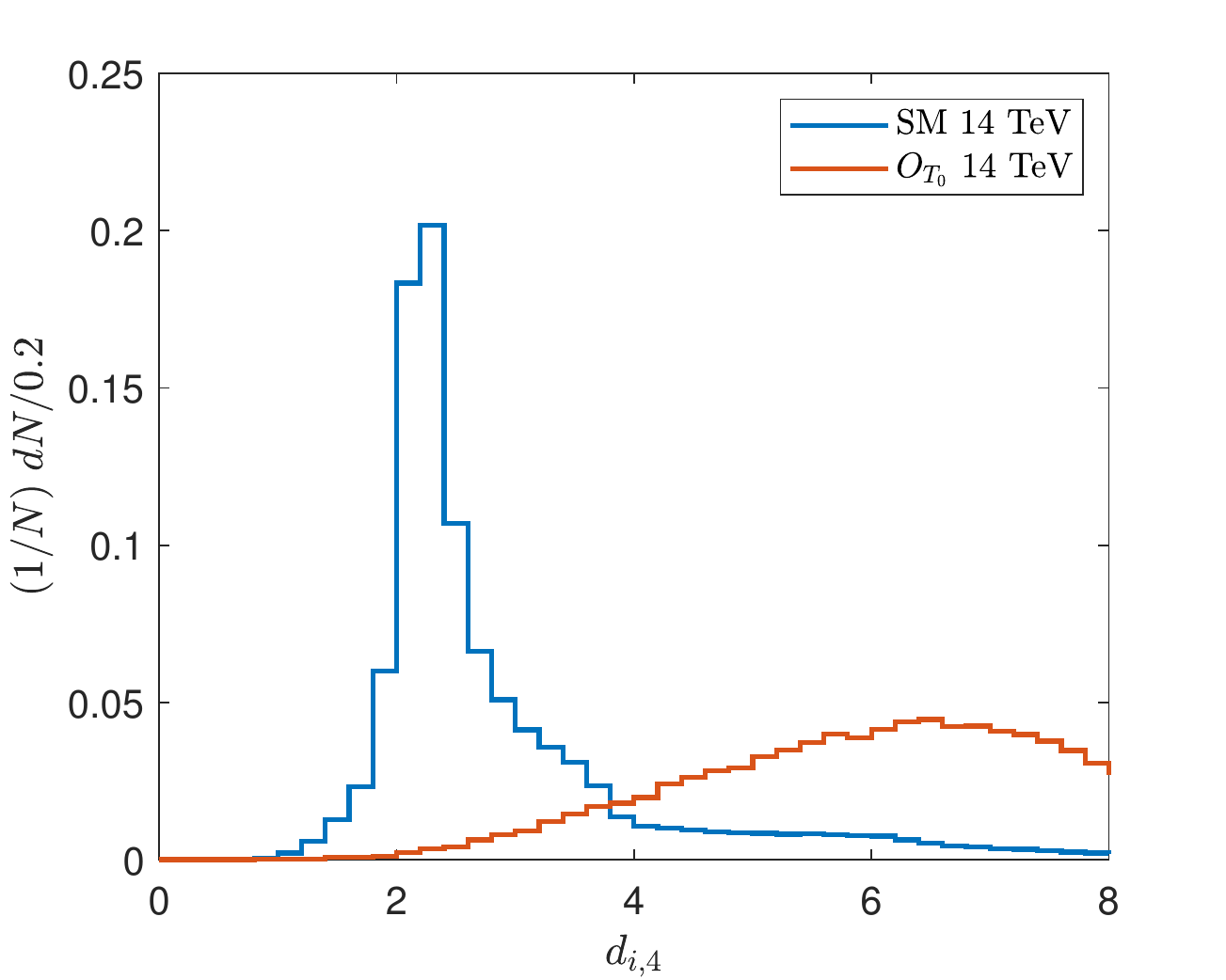}}
\resizebox{0.24\textwidth}{!}{\includegraphics{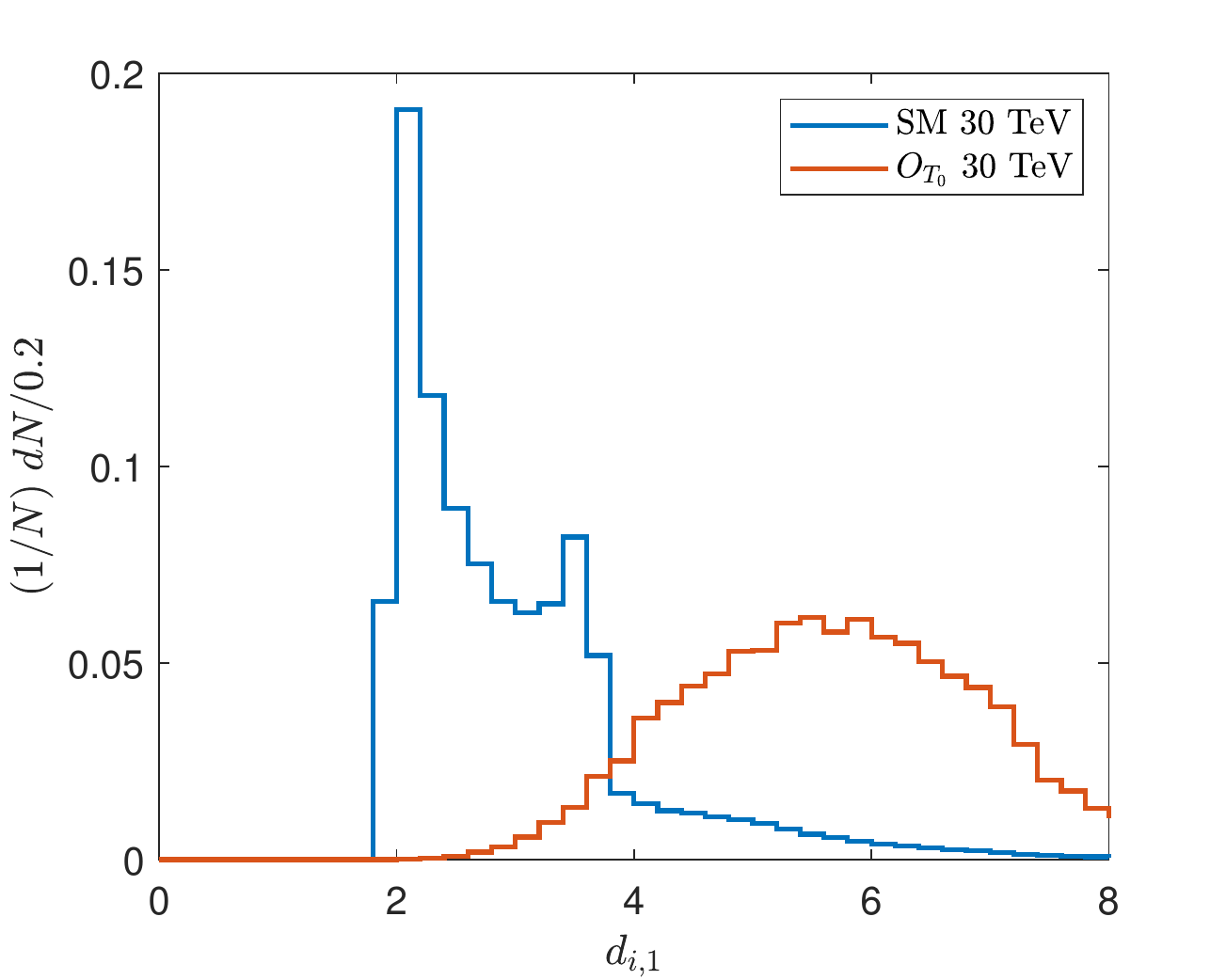}}
\resizebox{0.24\textwidth}{!}{\includegraphics{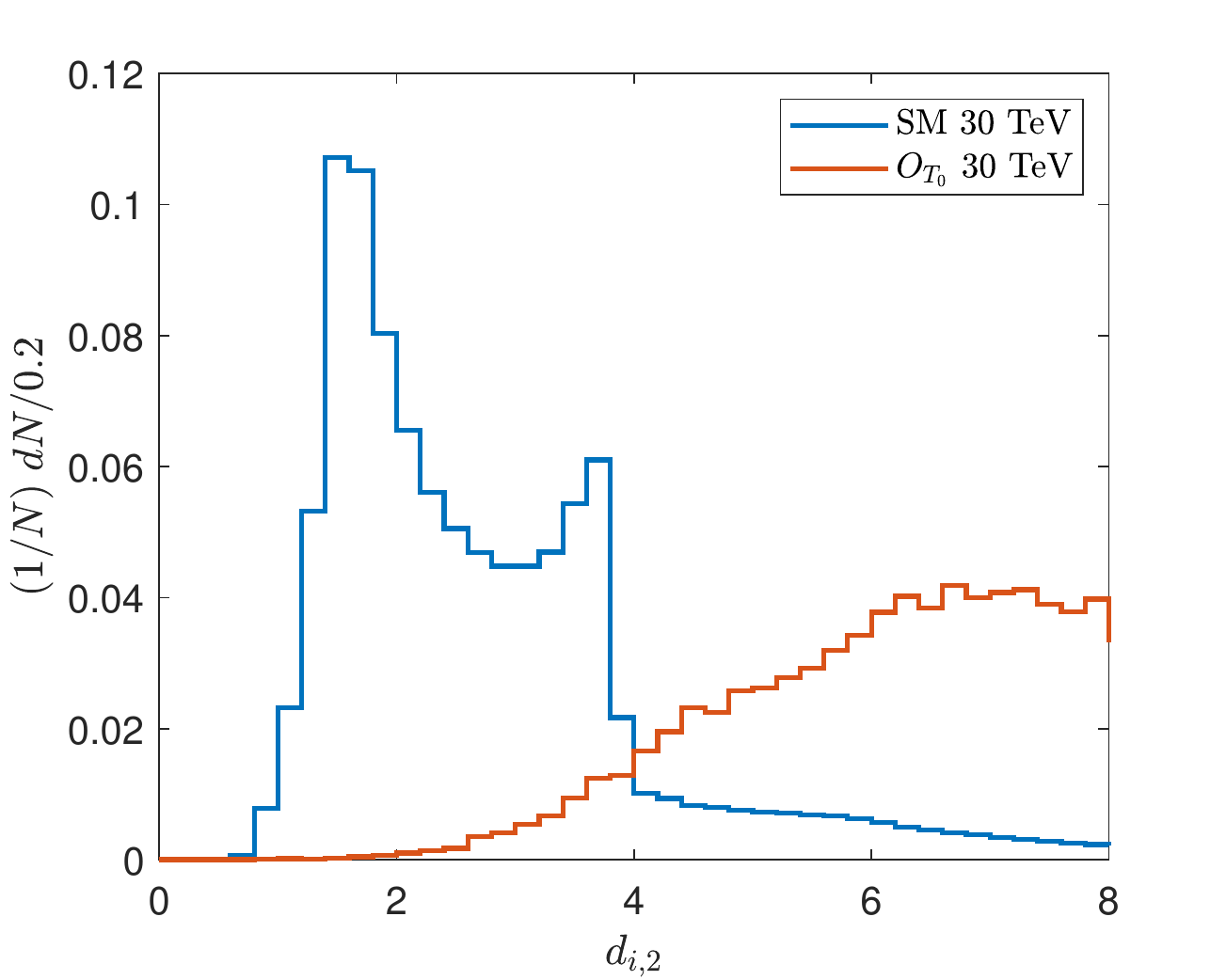}}
\resizebox{0.24\textwidth}{!}{\includegraphics{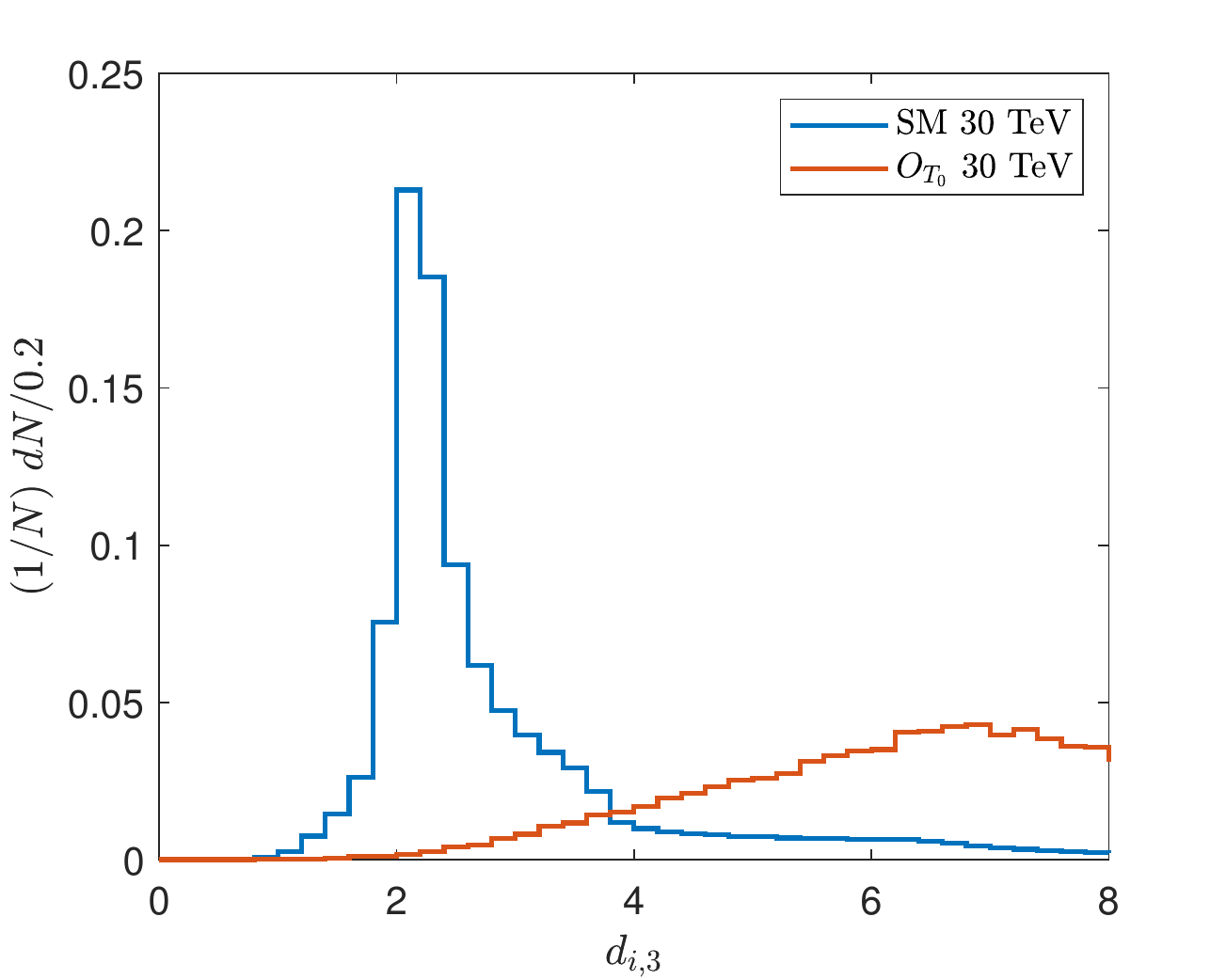}}
\resizebox{0.24\textwidth}{!}{\includegraphics{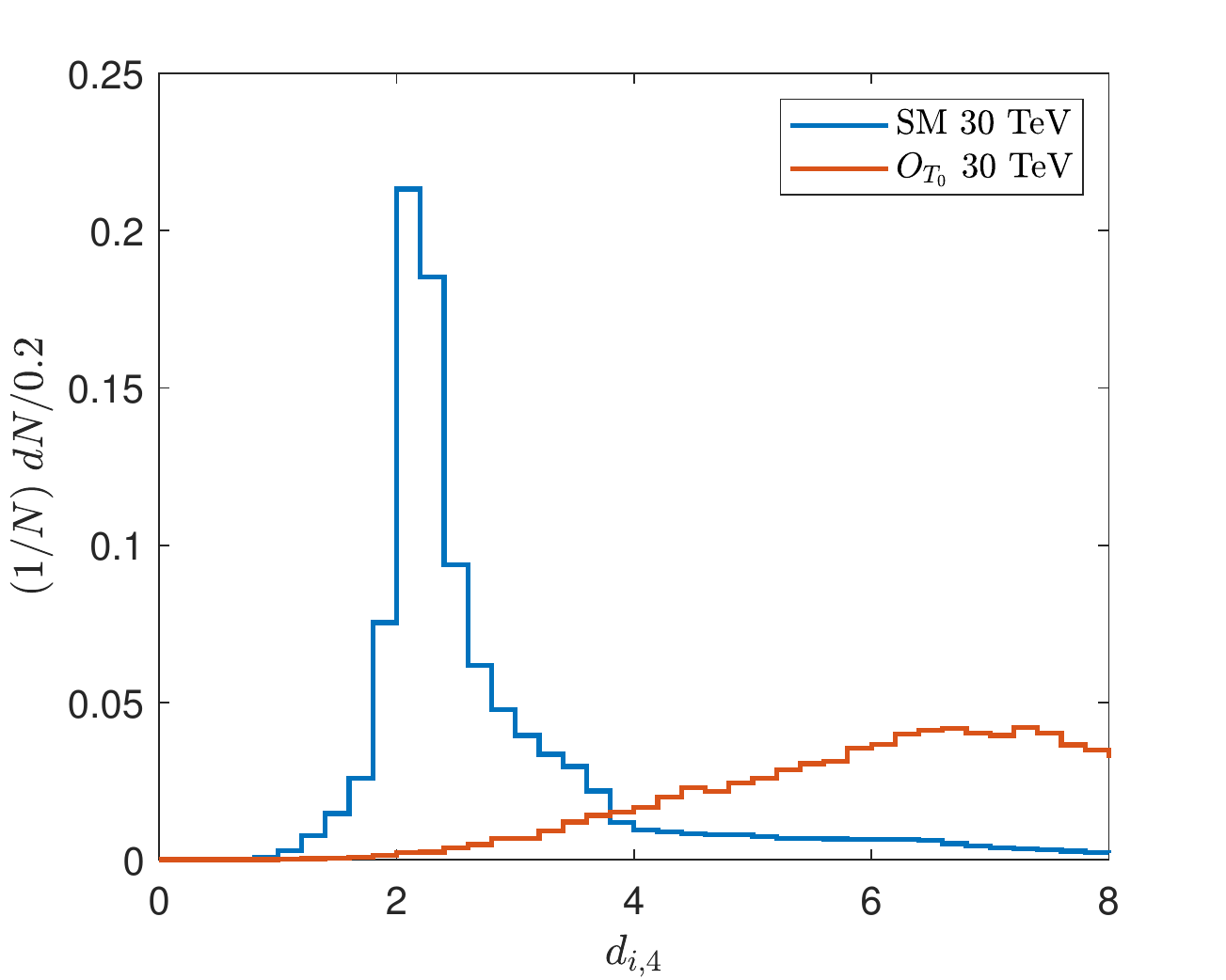}}
\caption{\label{fig:1500P3}Normalized distributions of $d_{i,1\leq j \leq 4}$ for the SM and $O_{T_0}$ at $\sqrt{s} =  3\;{\rm TeV}$~(the first row), $10\;{\rm TeV}$~(the second row), $14\;{\rm TeV}$~(the third row), and $30\;{\rm TeV}$~(the forth row).}
\end{center}
\end{figure*}

Taking the case of $O_{T_0}$ at $\sqrt{s} =  3$, $10$, $14$ and $30\;{\rm TeV}$ as examples, the standard deviations of $\tilde{x}_i^{\rm SM,j}$~(denoted as $\tilde{\epsilon}^{\rm SM,j}$) are shown in Fig.~\ref{fig:Variances}.
It can be shown that, $\tilde{\epsilon}^{\rm SM,j}$ with $1\leq j\leq 4$ are much larger than the other $\tilde{\epsilon}^{\rm SM,j}$'s, which indicates that the background points are mainly distributed along first $4$ eigenvectors.
Therefore, in this paper, we use $m'=4$.
We have also verified that, the improvements of expected constraints on the operator coefficients one can archive with a larger $m'$ are negligible compared with the case of $m'=4$.
The normalized distributions of $d_{i,j}^{\rm SM}$ and $d_{i,j}^{\rm tar}$ are shown in Fig.~\ref{fig:1500P3}.
It can be seen that, $d_{i,j}^{\rm tar}$ are generally larger than $d_{i,j}^{\rm SM}$ as expected.
In this paper, we use $d_{i,j}^{\rm tar}$ as the anomaly score to discriminate the signal from the background.
The means and standard deviations for $p^{\rm SM}$ used in z-score standardization are listed in Sec.~\ref{sec:ap1}, so as the components of $\vec{\eta} ^{\rm SM}_{1,2,3,4}$.

Although we can not care about the physical meanings of $d_{i,1}$, $d_{i,2}$, $d_{i,3}$ and $d_{i,4}$ when searching for NP signals using a AD approach such as PCAAD. 
But there is still an interesting and noteworthy phenomenon. 
For the SM, the positions of the $d_{i,1}$, $d_{i,2}$, $d_{i,3}$ and $d_{i,4}$ peaks are delayed to around 2, and the peak of $d_{i,2}$ is closer to $0$ compared with the others. 
Because of decentralization, the center points are those with high energies and small momenta, which cannot be the four momenta of photons which should be light-like. 
This is the reason that $d_{i,1}^{\rm SM}$ cannot be too close to $0$.
And $\vec{\eta} _2^{\rm SM}$ approximately corresponds to the directions of all photon momenta along the $\bf z$-axis. 
Thus $d_{i,2}$ distributed at a position more towards $0$ shows that the photons are more inclined to the $\bf z$-axis, which can be seen as a consequence of infrared divergence. 
Meanwhile, for $d_{i,2}$ which also peaks at a position larger than $0$ instead of $0$ due to the cuts to avoid infrared divergence.

Another noteworthy result is that $\tilde{\epsilon} ^{\rm SM,9,10,11,12}\approx 0$.
The PCA can automatically find out the redundant information.
Ignoring the effect of detector simulation, $4$ of the $12$ variables are linearly related to the others by $\sum p_n^{\gamma} = (\sqrt{s},0,0,0)$ where $p_{n=1,2,3}^{\gamma}$ are momenta of the photons, and there are $3$ other nonlinear relations $(p_n^{\gamma})^2=0$.
The numerical results of $\vec{\eta} ^{\rm SM}_{9,10,11,12}$ at $\sqrt{s}=3\;{\rm TeV}$ indicates that,
\begin{equation}
\begin{split}
&\tilde{x}_i^9 \approx 0.0033\;{\rm GeV}^{-1}\left((E_1-1.452\times 10^3\;{\rm GeV})\right.\\
&\left.+(E_2-1.313\times 10^3\;{\rm GeV})+(E_3-2.351\times 10^2\;{\rm GeV} )\right)\\
&\tilde{x}_i^{10}\approx 10^{-3}\;{\rm GeV}^{-1}\left(0.67p_1^x-0.98p_1^y\right.\\
&\left.+0.67p_2^x-0.98p_2^y+0.67p_3^x-0.98p_3^y\right)\\
&\tilde{x}_i^{11}\approx 10^{-3}\;{\rm GeV}^{-1}\left(0.98p_1^x+0.67p_1^y\right.\\
&\left.+0.98p_2^x+0.67p_2^y+0.98p_3^x+0.67p_3^y\right)\\
&\tilde{x}_i^{12}\approx 0.00062\;{\rm GeV}^{-1}\left(p_1^z+p_2^z+p_3^z\right)\\
\end{split}
\label{eq.xtilde}
\end{equation}
where $E_{1,2,3}$ are energies of the three photons, $p_{1,2,3}^{x,y,z}$ are the components of $p_n^{\gamma}$.
$\tilde{\epsilon} ^{\rm SM}_{9,10,11,12}\approx 0$ indicate that $\tilde{x}_{9,10,11,12}^{\rm SM}$ are almost constants.
This corresponds to $\sum p_n^{\gamma} = (\sqrt{s},0,0,0)$.
However, to cleanly remove the non-linear redundant variables, non-linear PCA must be used~\cite{nonlinearpca}.

\section{\label{sec4}Constraints on the coefficients}

\begin{table}[htbp]
\centering
\begin{tabular}{c|c|c} 
\hline
 $\sqrt{s}$ & 3\;{\rm TeV} &10\;{\rm TeV} \\ 
 \hline
 Unit of coefficient & $ ({\rm TeV}^{-4})$ & $({\rm TeV}^{-4})$ \\
\hline
$\left| f_{T_0}/\Lambda ^4\right|$ & $\leq 0.3$&$\leq 0.0015$ \\ 
\hline
$\left| f_{T_2}/\Lambda ^4\right|$ & $\leq 0.5$&$\leq 0.002$ \\
\hline
$\left| f_{T_5}/\Lambda ^4\right|$ & $\leq 0.06$ &$\leq 0.0003$ \\
\hline
$\left| f_{T_7}/\Lambda ^4\right|$ & $\leq 0.1$ & $\leq 0.0005$ \\
\hline
$\left| f_{T_8}/\Lambda ^4\right|$ & $\leq 0.01$& $\leq 0.00005$ \\ 
\hline
$\left| f_{T_9}/\Lambda ^4\right|$ & $\leq 0.016$& $\leq 0.00008$ \\  
\hline
$\sqrt{s}$ & $14\;{\rm TeV}$ & $30\;{\rm TeV}$ \\ 
\hline
Unit of coefficient &$({\rm TeV}^{-4})$&$({\rm TeV}^{-4})$\\
\hline
$\left|f_{T_0}/\Lambda ^4\right|$ &$\leq 0.0005$& $\leq 0.00005$\\ 
\hline
$\left|f_{T_2}/\Lambda ^4\right|$ & $\leq 0.0008$& $\leq 0.00008$\\
\hline
$\left|f_{T_5}/\Lambda ^4\right|$ & $\leq 0.0001$& $\leq 0.000008$\\
\hline
$\left|f_{T_7}/\Lambda ^4\right|$ &$\leq 0.00015$ &$\leq 0.000015$ \\
\hline
$\left|f_{T_8}/\Lambda ^4\right|$ & $\leq 0.000015$ & $\leq 0.0000015$ \\ 
\hline
$\left|f_{T_9}/\Lambda ^4 \right|$ & $\leq 0.00002$ & $\leq 0.000002$ \\  
\hline
\end{tabular}
\caption{The ranges of operator coefficients used in the scanning}
\label{table:coefficientscan}
\end{table}

When no NP signal is found, the PCAAD event selection strategy can also be used to constrain the coefficients of NP. 
To this end, we generate events with the coefficients in Table ~\ref{table:coefficientscan}.
In this section, the target data-sets consist of the events generated with the SM, NP, and interference between the SM and NP included. 
In Ref.~\cite{triphoton}, $p_{T,\gamma} > 0.12 E_{\rm beam}$ is used as a part of the event selection strategy, where $E_{\rm beam}$ is the energy of the beam. To avoid dealing with too many events, when generating events, the standard cut requires $p_{T,\gamma} > 0.1 E_{\rm beam}$ while the other standard cuts are the same as those in Eq.~(\ref{eq.standardcuts}). 

\begin{figure*}[htbp]
\begin{center}
\resizebox{0.32\textwidth}{!}{\includegraphics{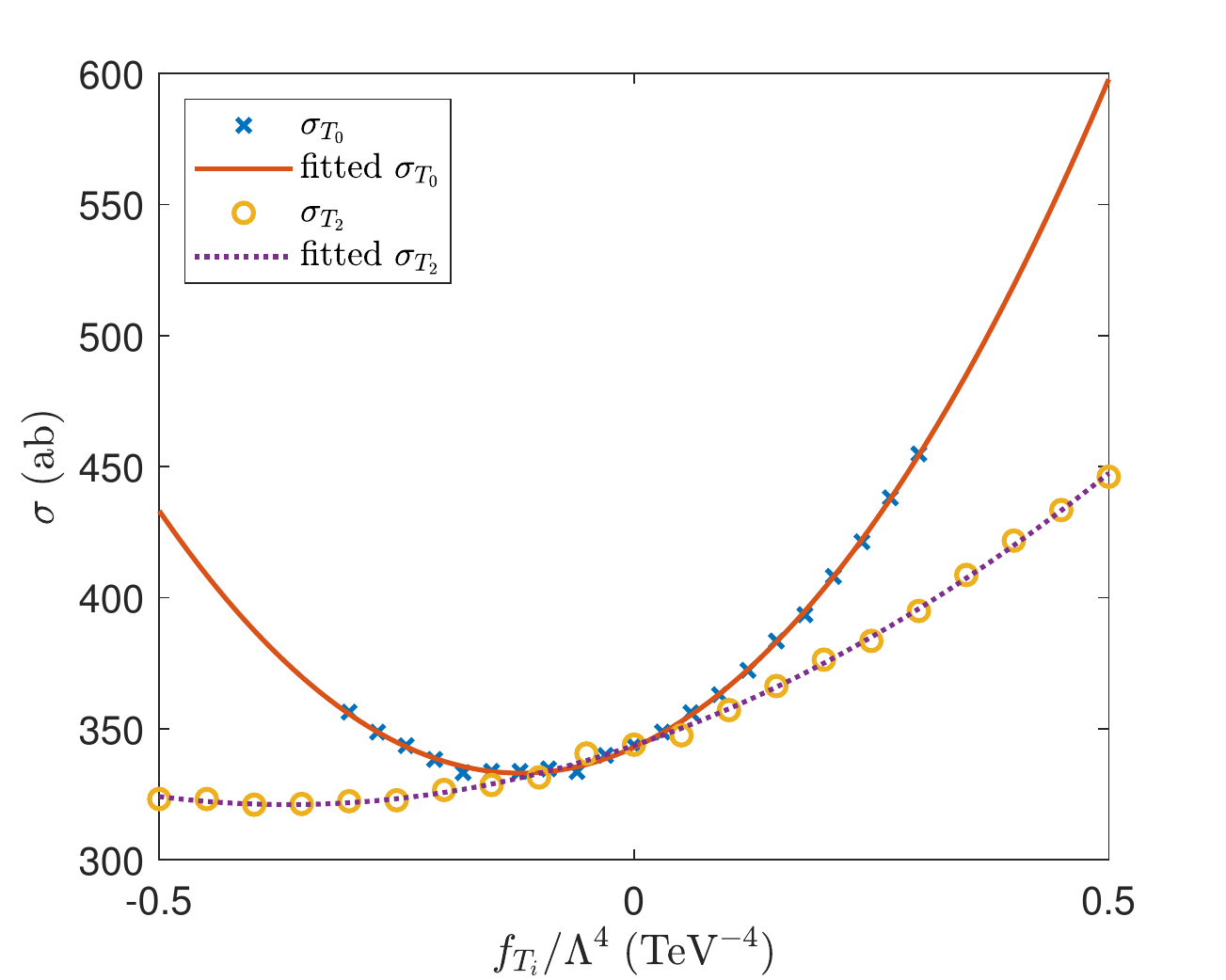}}
\resizebox{0.32\textwidth}{!}{\includegraphics{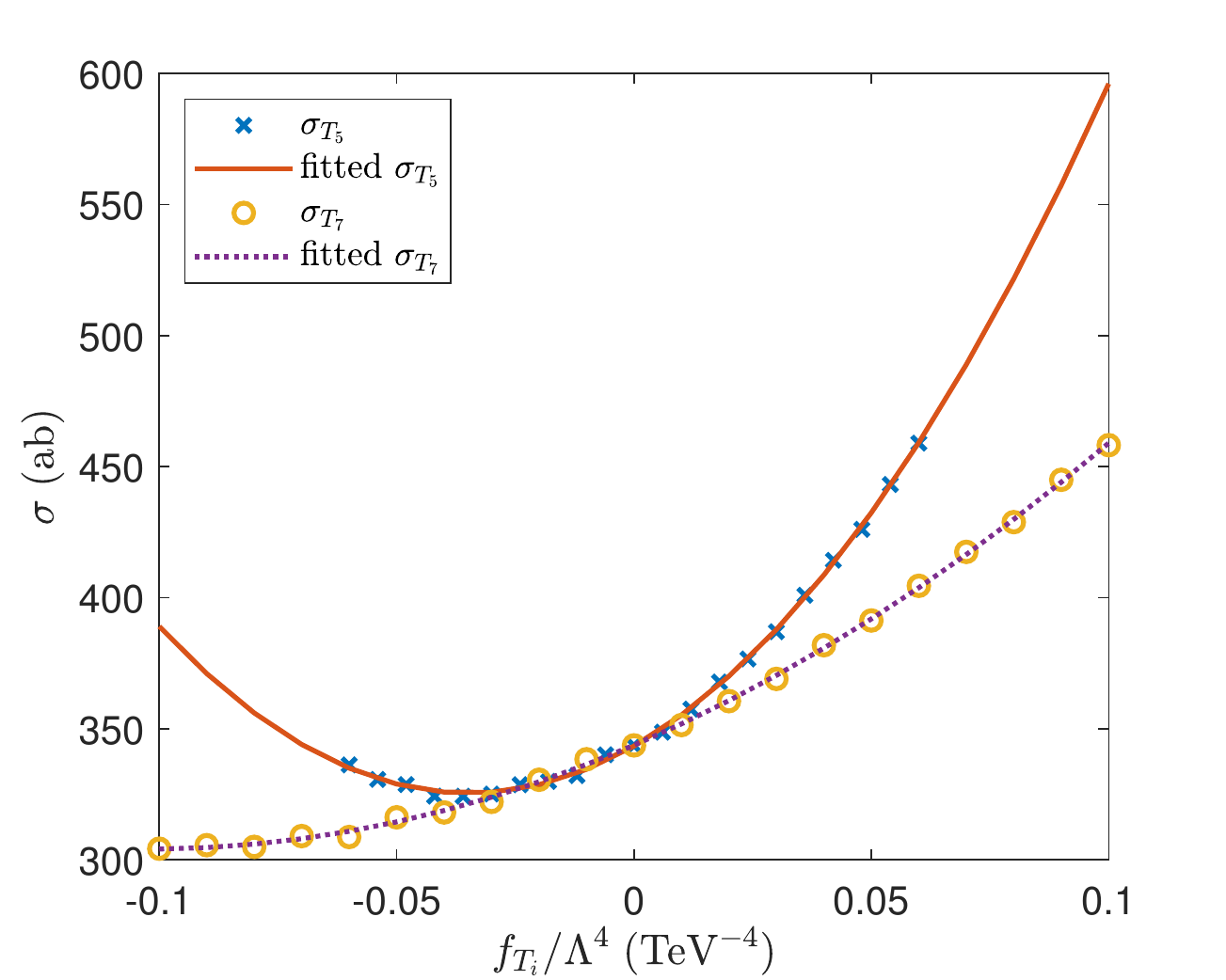}}
\resizebox{0.32\textwidth}{!}{\includegraphics{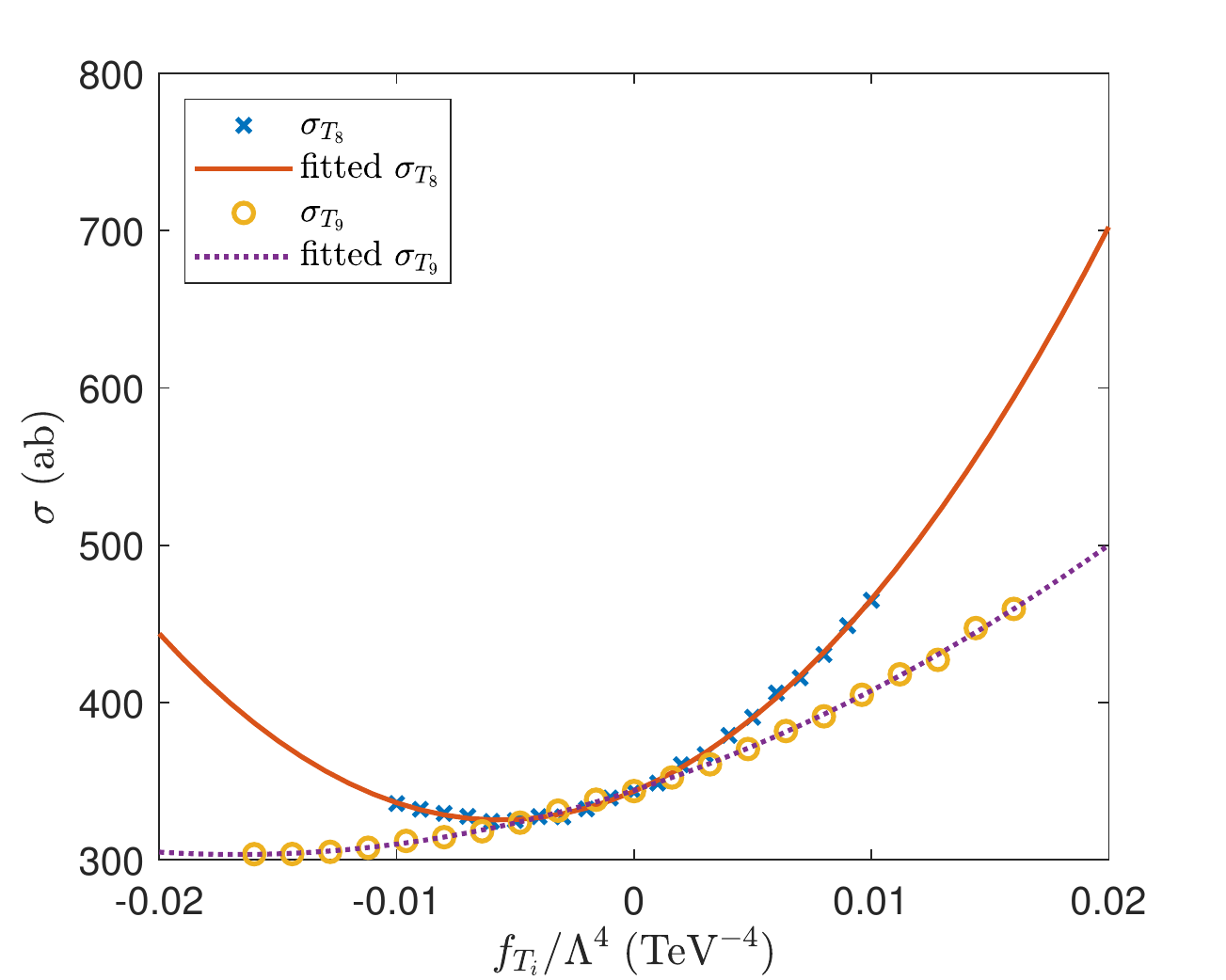}}
\resizebox{0.32\textwidth}{!}{\includegraphics{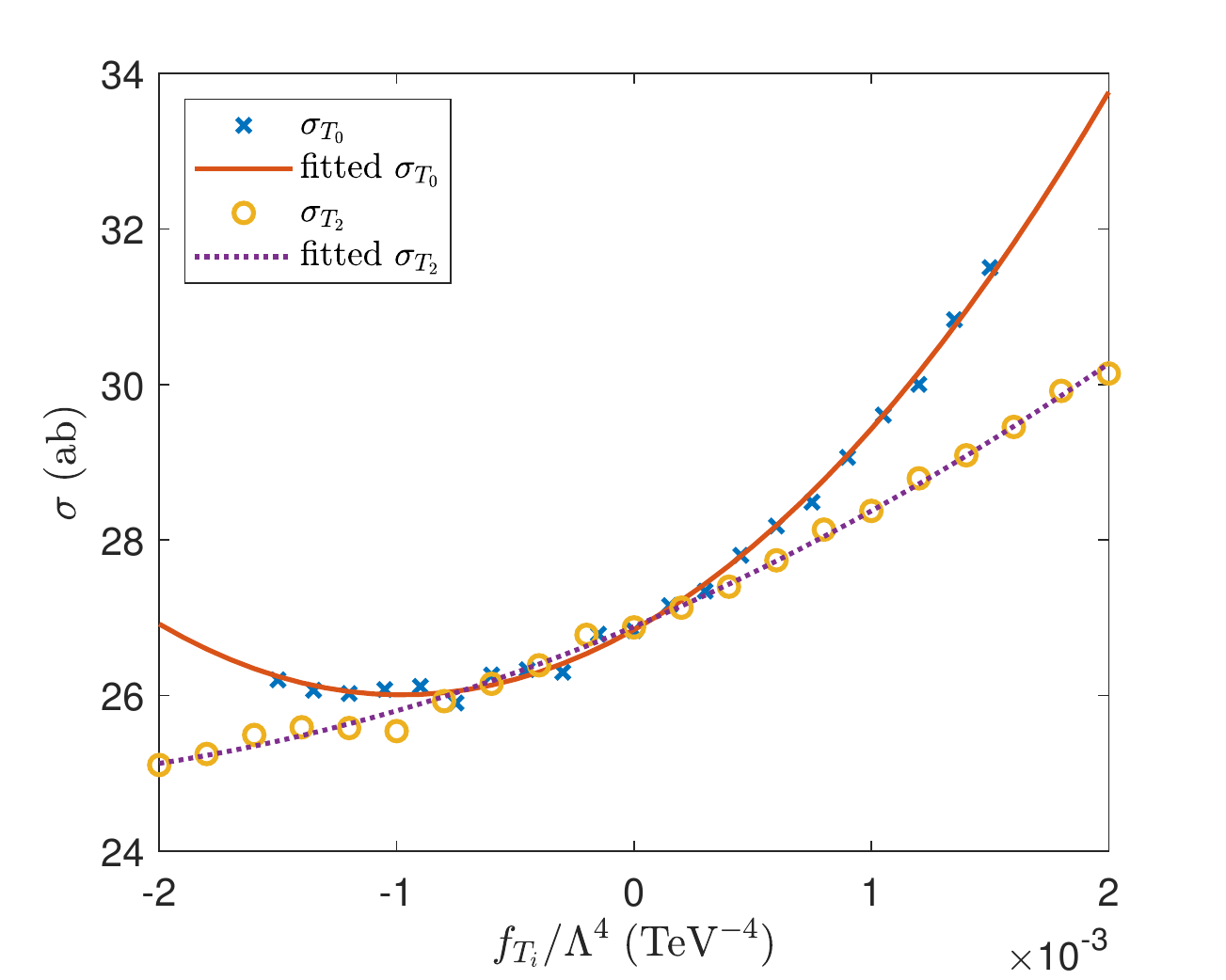}}
\resizebox{0.32\textwidth}{!}{\includegraphics{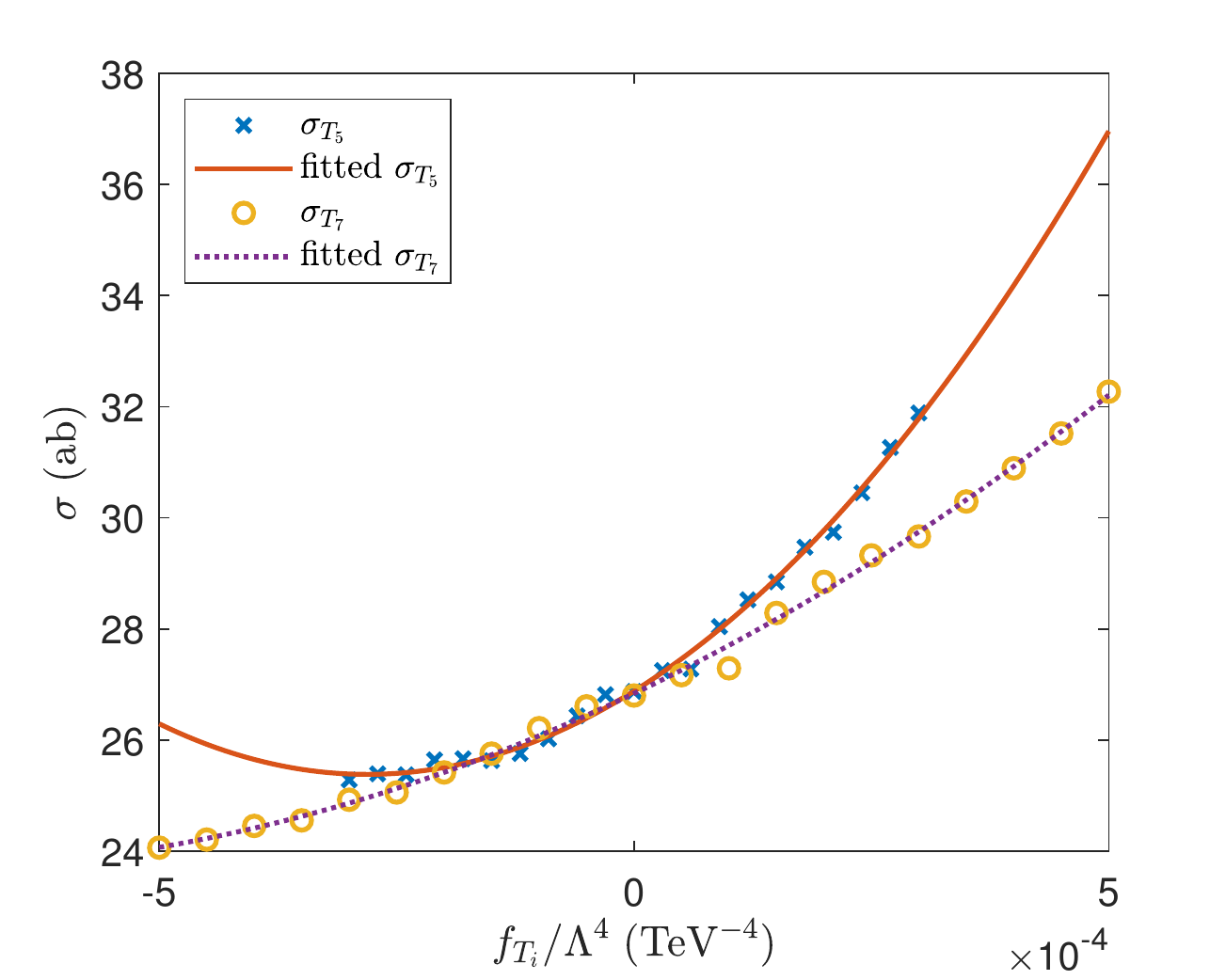}}
\resizebox{0.32\textwidth}{!}{\includegraphics{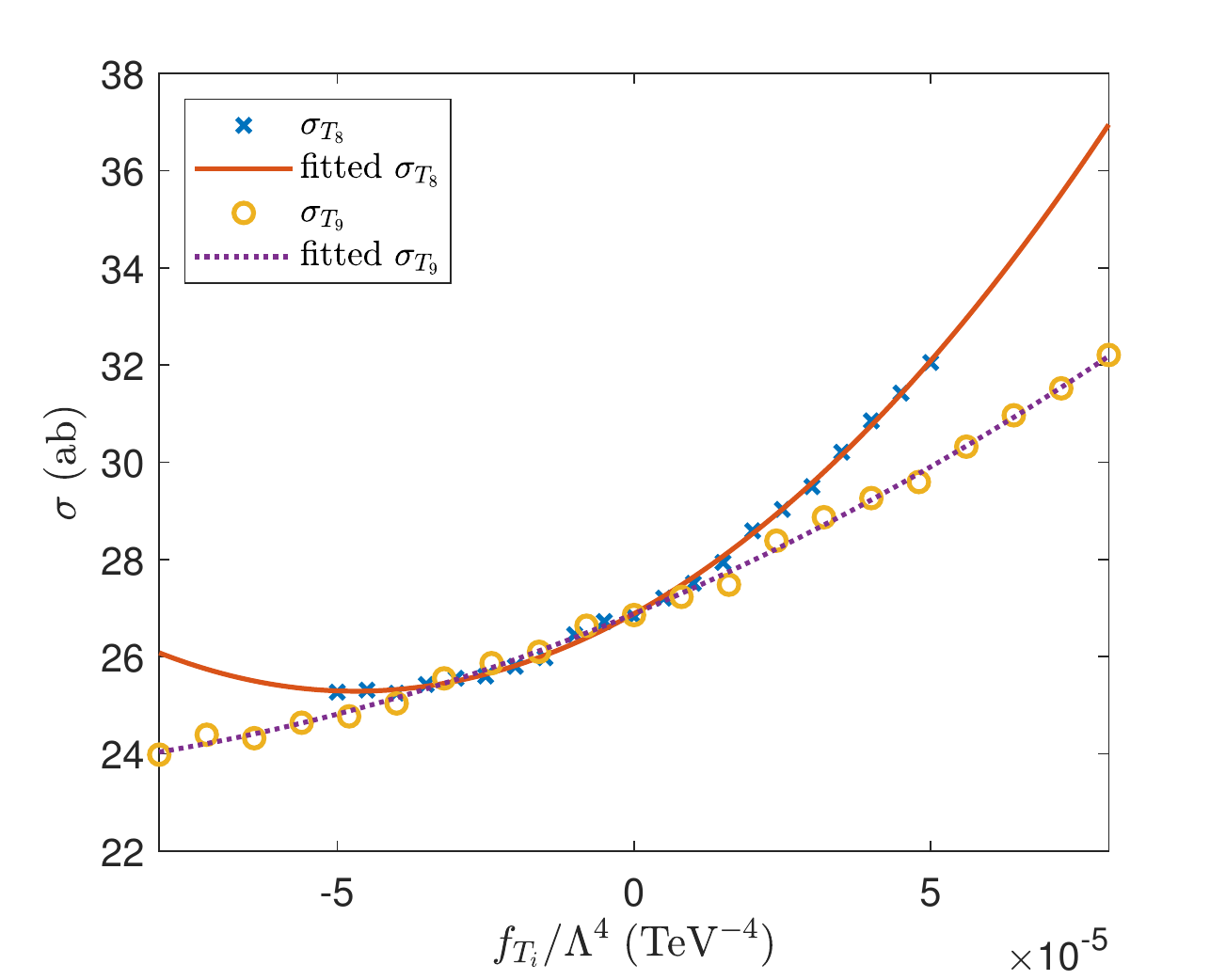}}
\resizebox{0.32\textwidth}{!}{\includegraphics{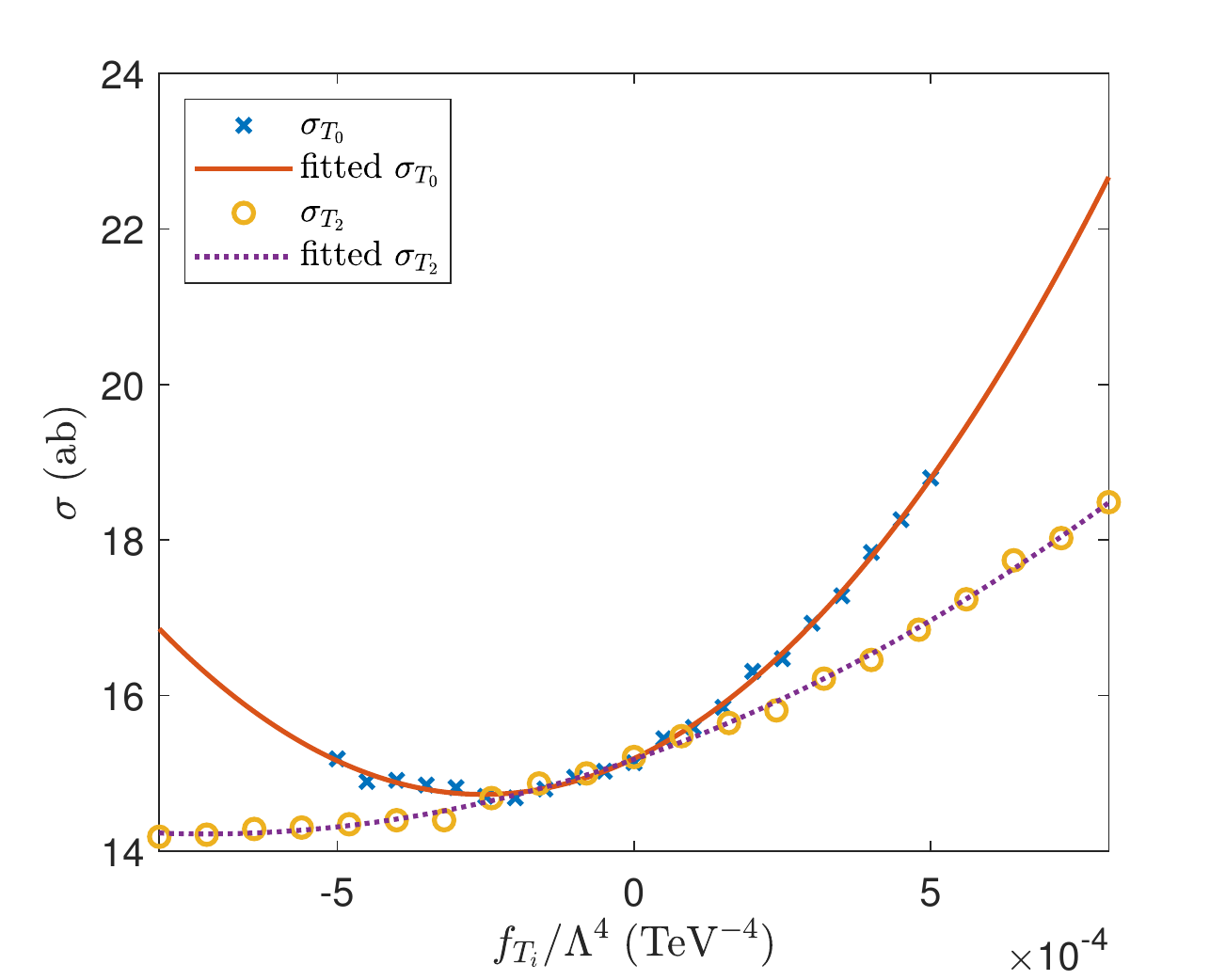}}
\resizebox{0.32\textwidth}{!}{\includegraphics{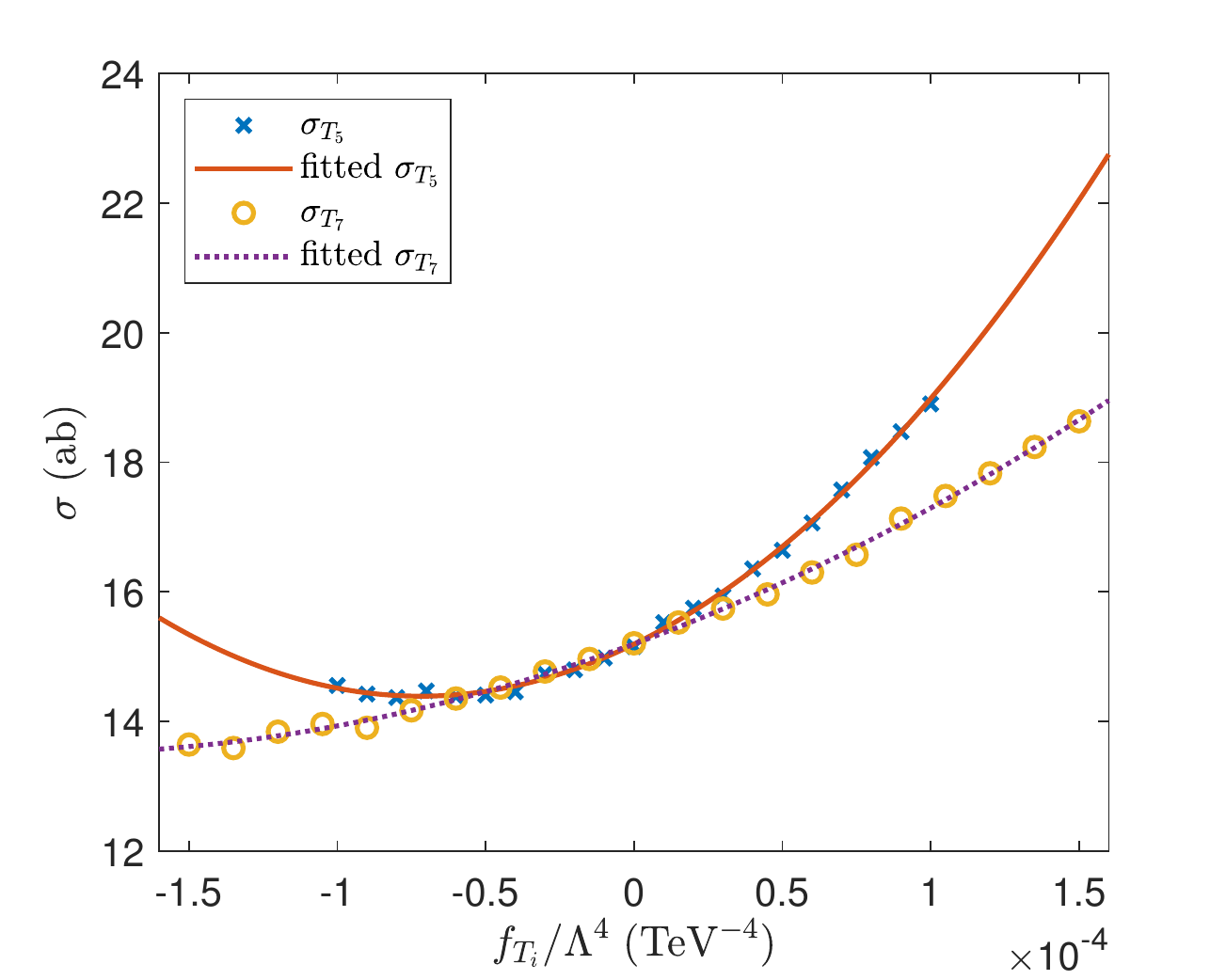}}
\resizebox{0.32\textwidth}{!}{\includegraphics{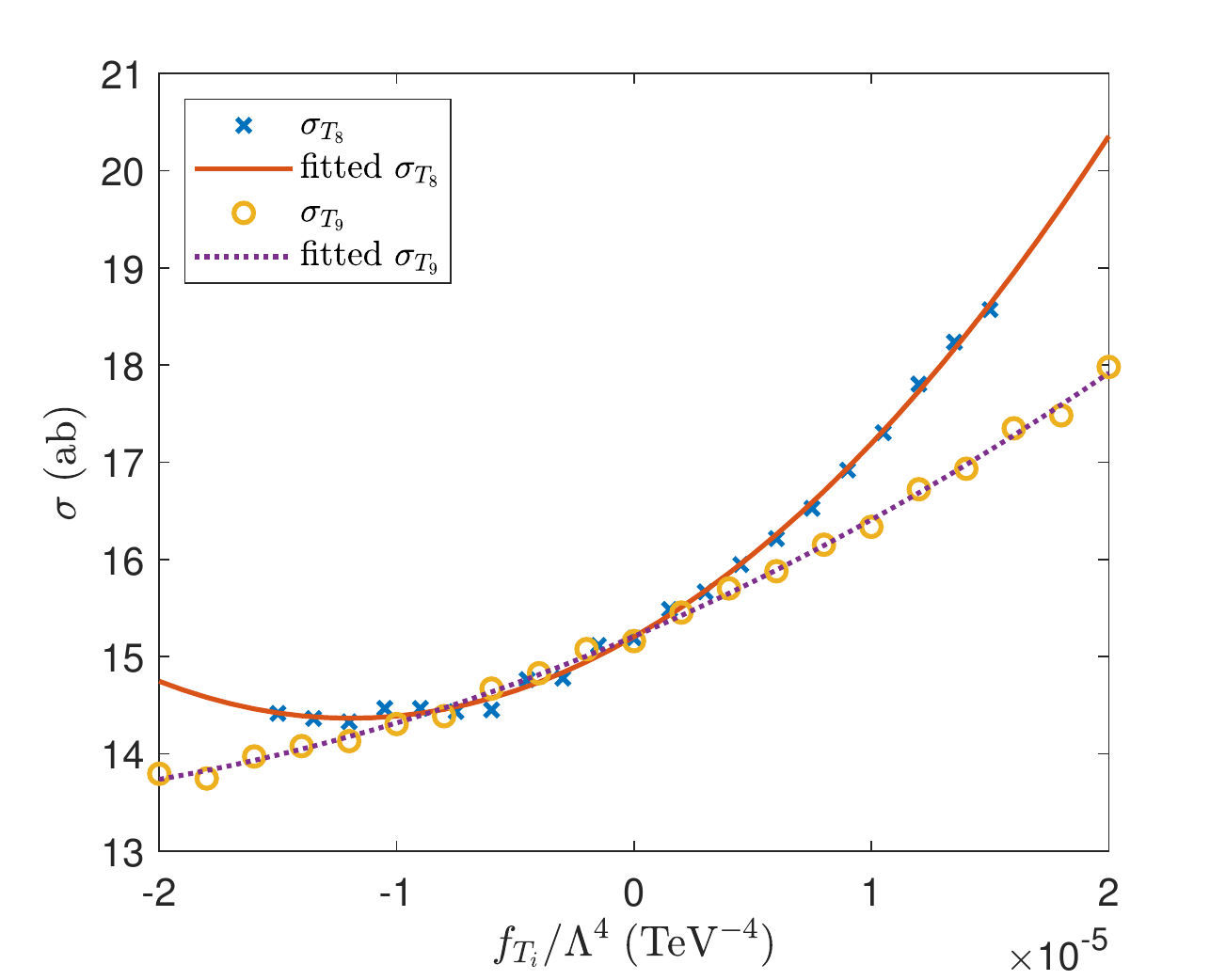}}
\resizebox{0.32\textwidth}{!}{\includegraphics{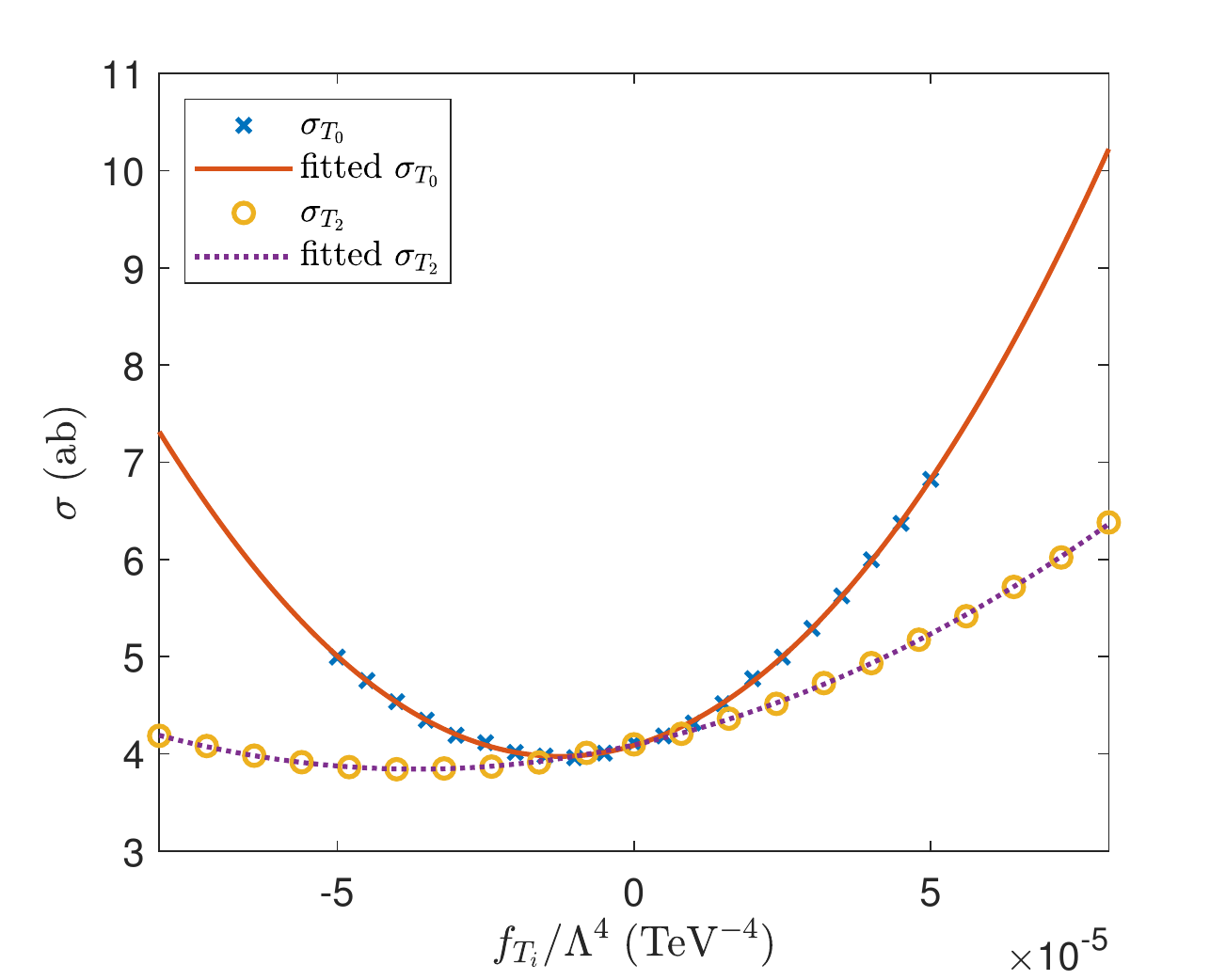}}
\resizebox{0.32\textwidth}{!}{\includegraphics{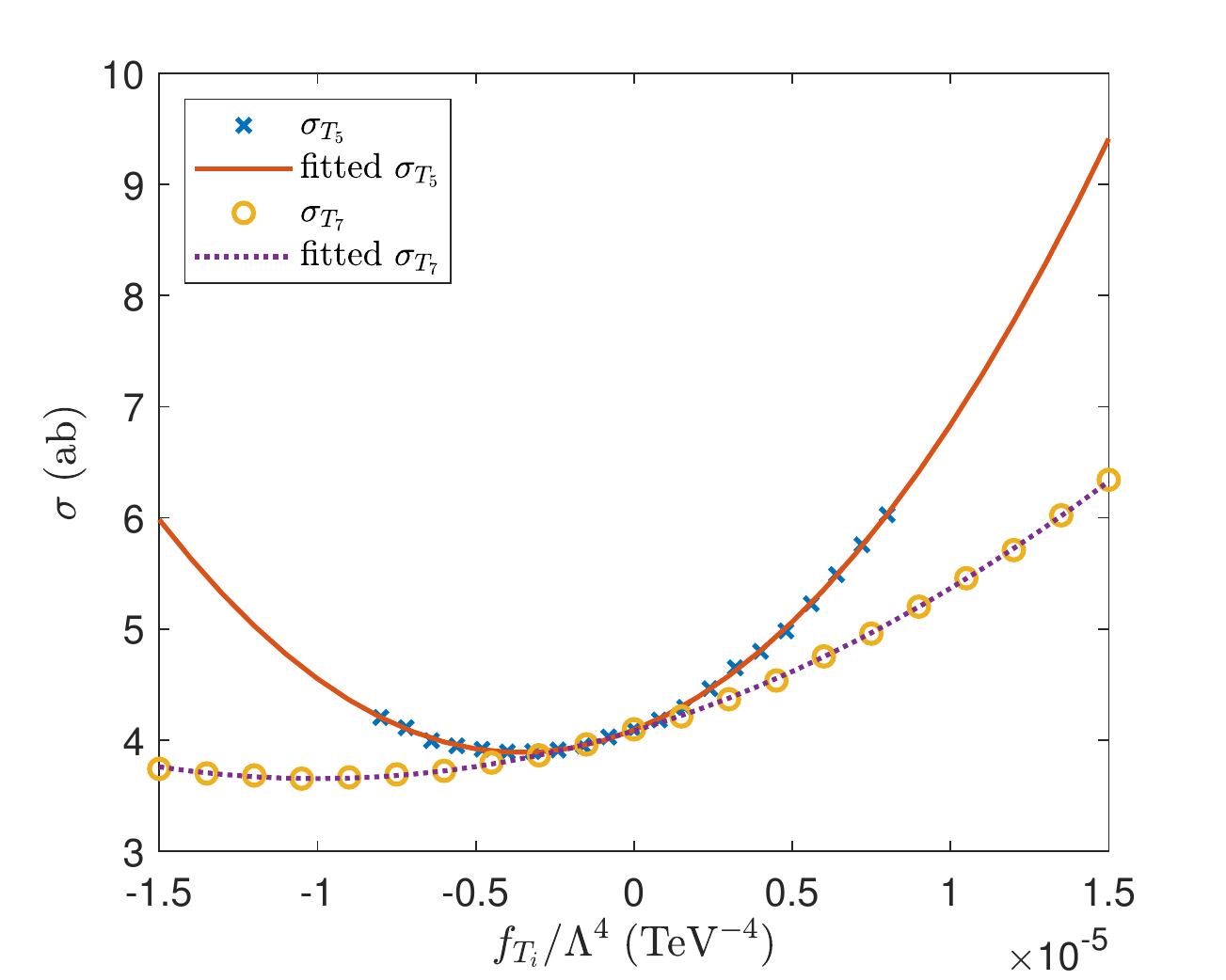}}
\resizebox{0.32\textwidth}{!}{\includegraphics{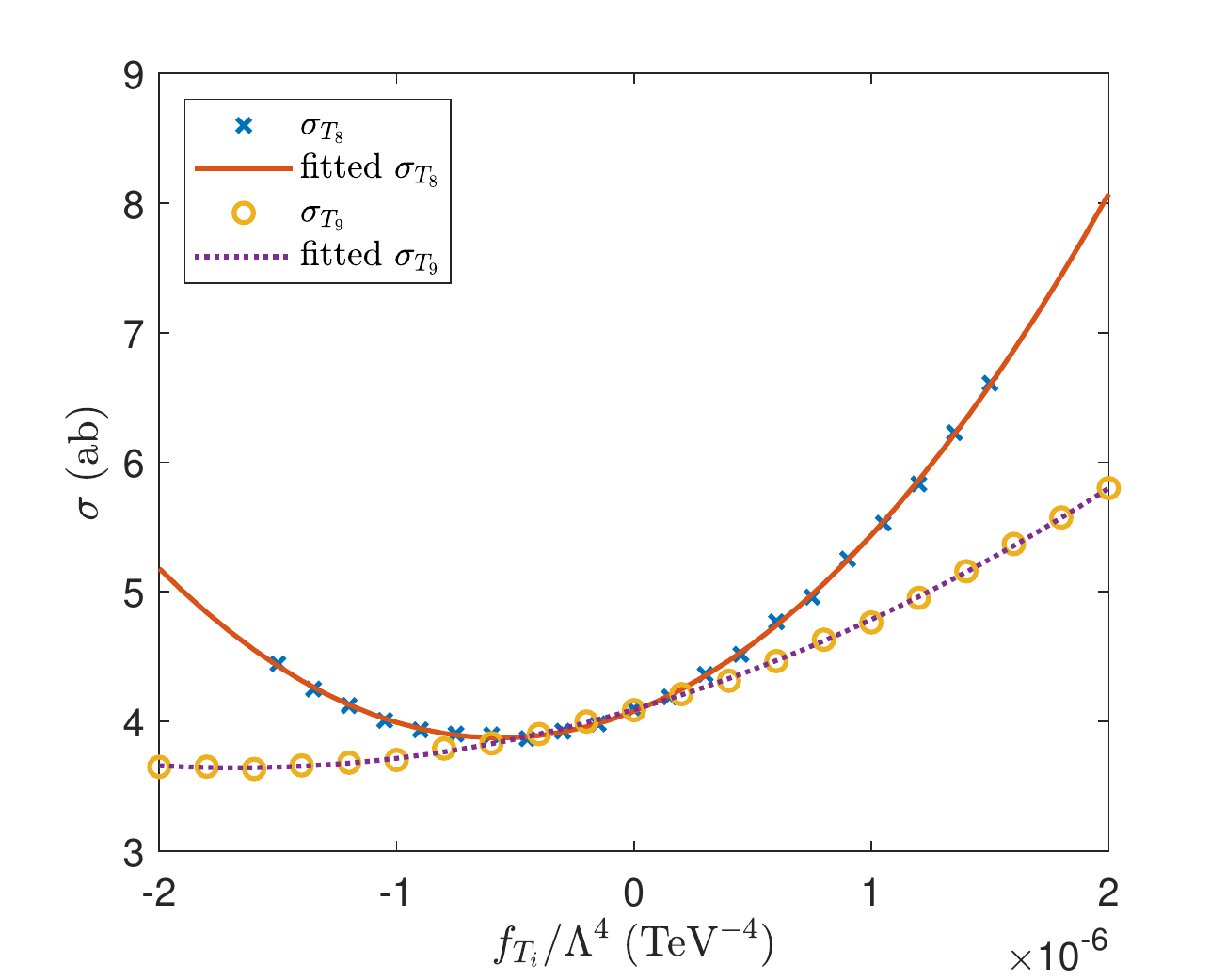}}
\caption{\label{fig:15000t8t9}
The cross-sections after cut and the fitted cross-sections at $\sqrt{s}$ =  3\;{\rm TeV}~(row 1), 10\;{\rm TeV}~(row 2), 14\;{\rm TeV}~(row 3), and 30\;{\rm TeV}~(row 4).}
\end{center}
\end{figure*}

The cross-section can be expressed as a parabola of $f_{T_{i}}/\Lambda^{4}$, as $\sigma=\sigma_{\rm SM}+ \sigma_{\rm int} f_{T_{i}}/\Lambda^{4}+ \sigma_{\rm NP} \left(f_{T_{i}}/\Lambda^{4}\right)^2$, where $\sigma _{\rm SM}$ denotes the contribution of the SM, $f_{T_{i}}/\Lambda^{4} \sigma _{\rm int}$ denotes the interference between the SM and aQGCs, and $\sigma _{\rm NP} \left(f_{T_{i}}/\Lambda^{4}\right)^2$ is the contribution induced by aQGCs. 
After obtaining the cross section after cuts corresponding to each coefficient, we fitted the cross sections according to parabola using the least squares method.
For simplicity we use a same criterion for $d_{i,1\leq j \leq 4}$ for each $\sqrt{s}$~(denoted as $d$).
Taking $d_{i,1\leq j \leq 4}<4.2$ at $\sqrt{s}=3\;{\rm TeV}$ and $d_{i,1\leq j \leq 4}<4.8$ at $\sqrt{s}=10$, $14$, and $30\;{\rm TeV}$ as examples, the cross-sections after the event selection strategy and the fitted cross-sections are shown in Fig.~\ref{fig:15000t8t9}. 
It can be seen that the cross-sections also fit the parabola functions well after cuts.

\begin{figure}[htbp]
\begin{center}
\resizebox{0.48\hsize}{!}{\includegraphics{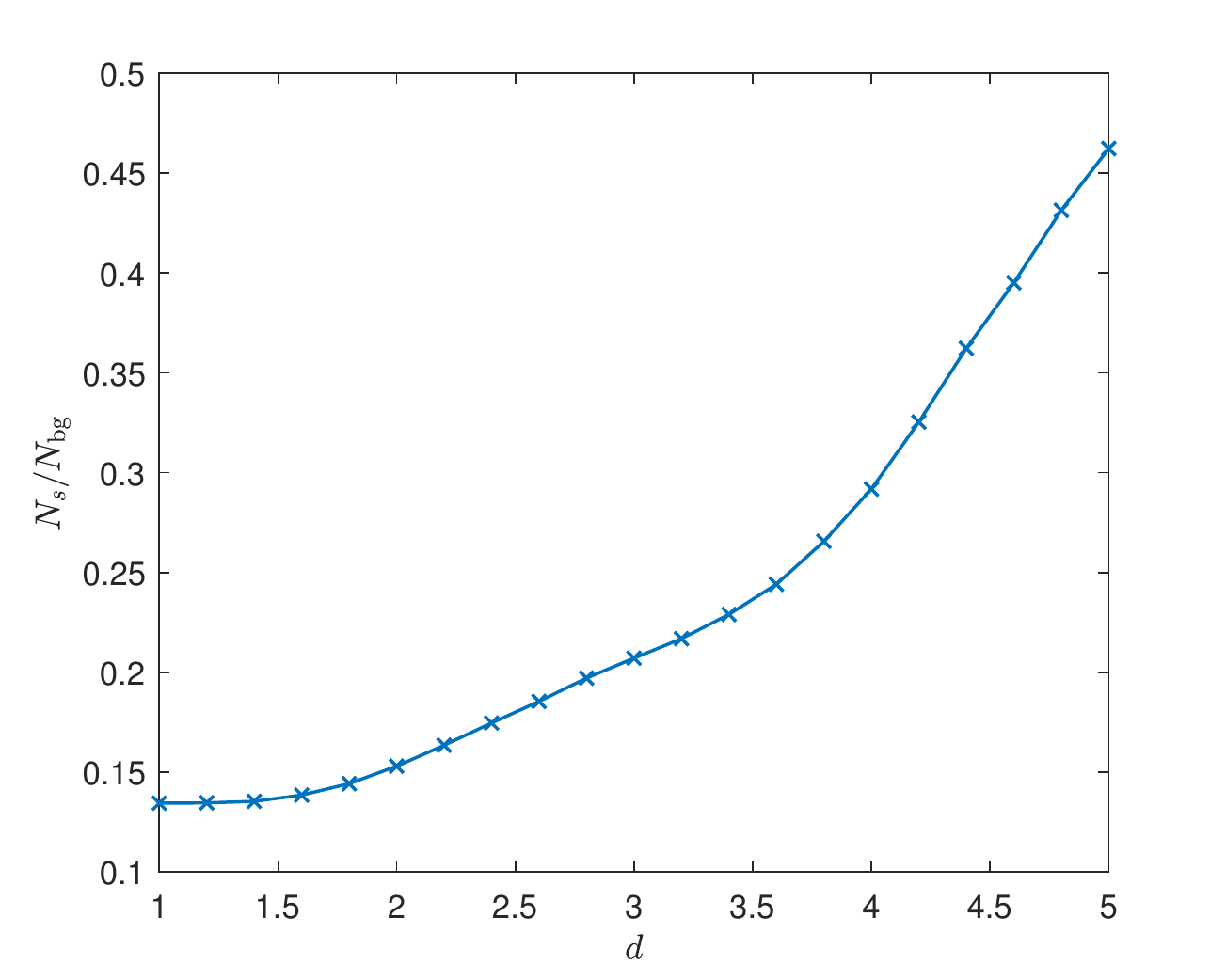}}
\resizebox{0.48\hsize}{!}{\includegraphics{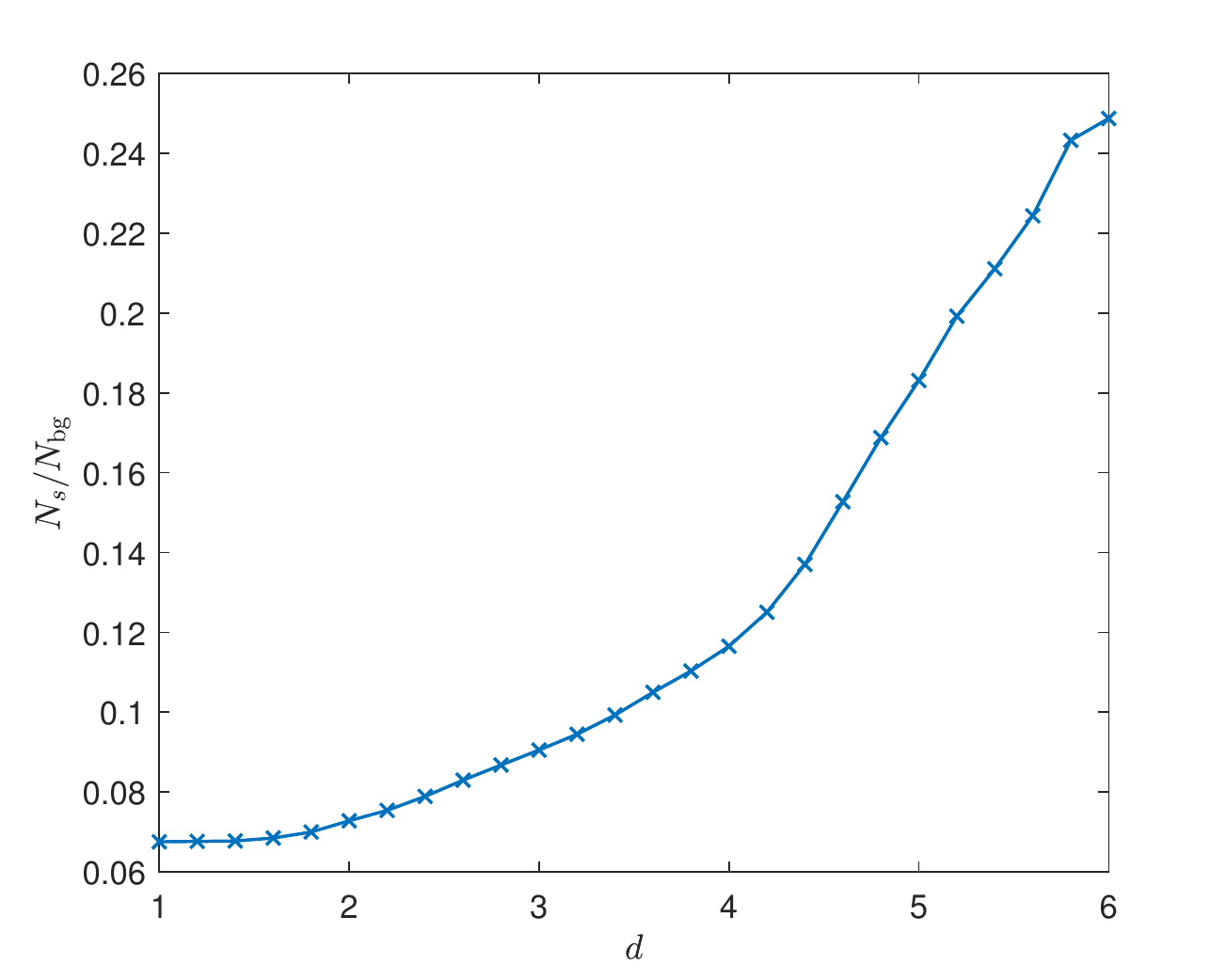}}\\
\resizebox{0.48\hsize}{!}{\includegraphics{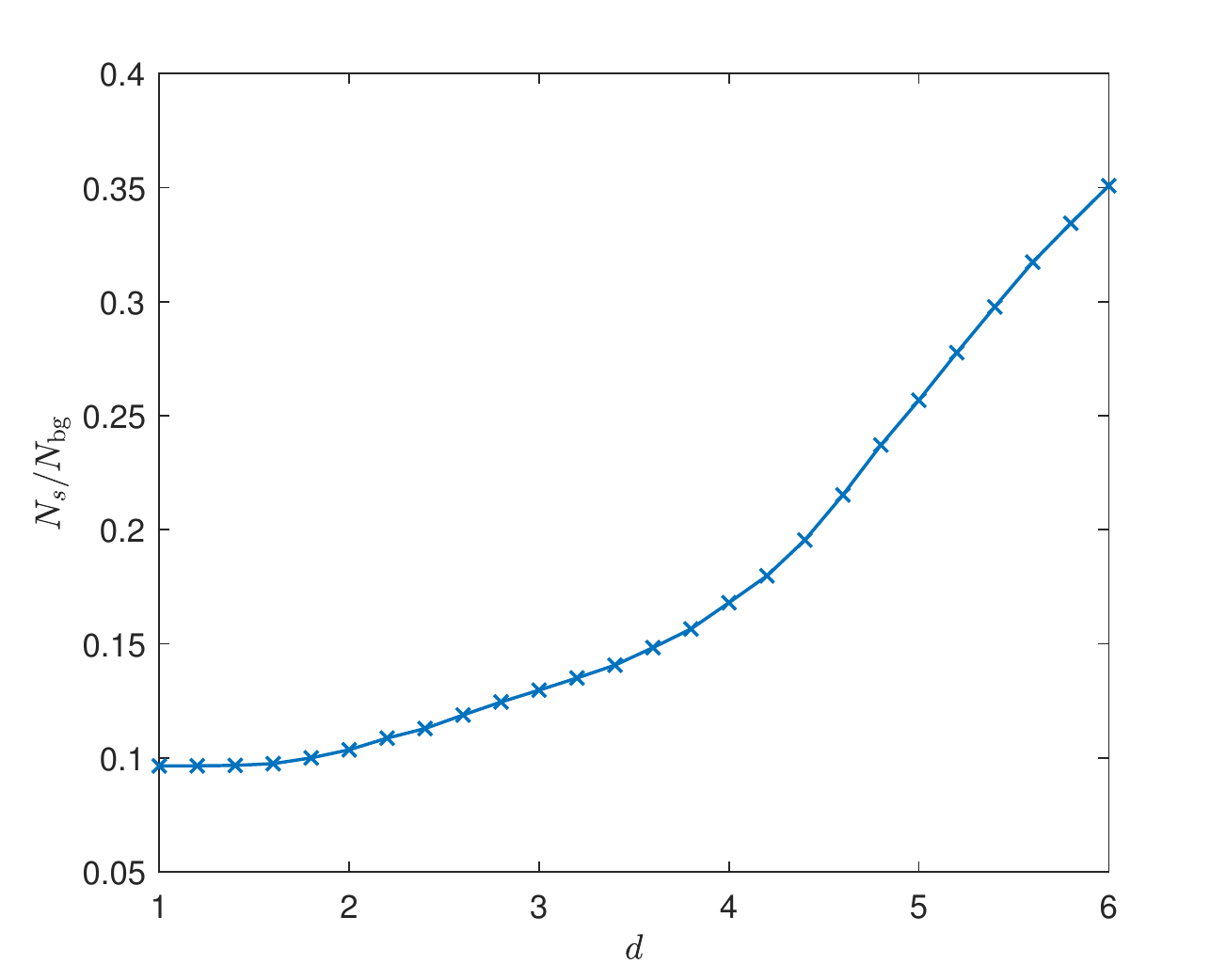}}
\resizebox{0.48\hsize}{!}{\includegraphics{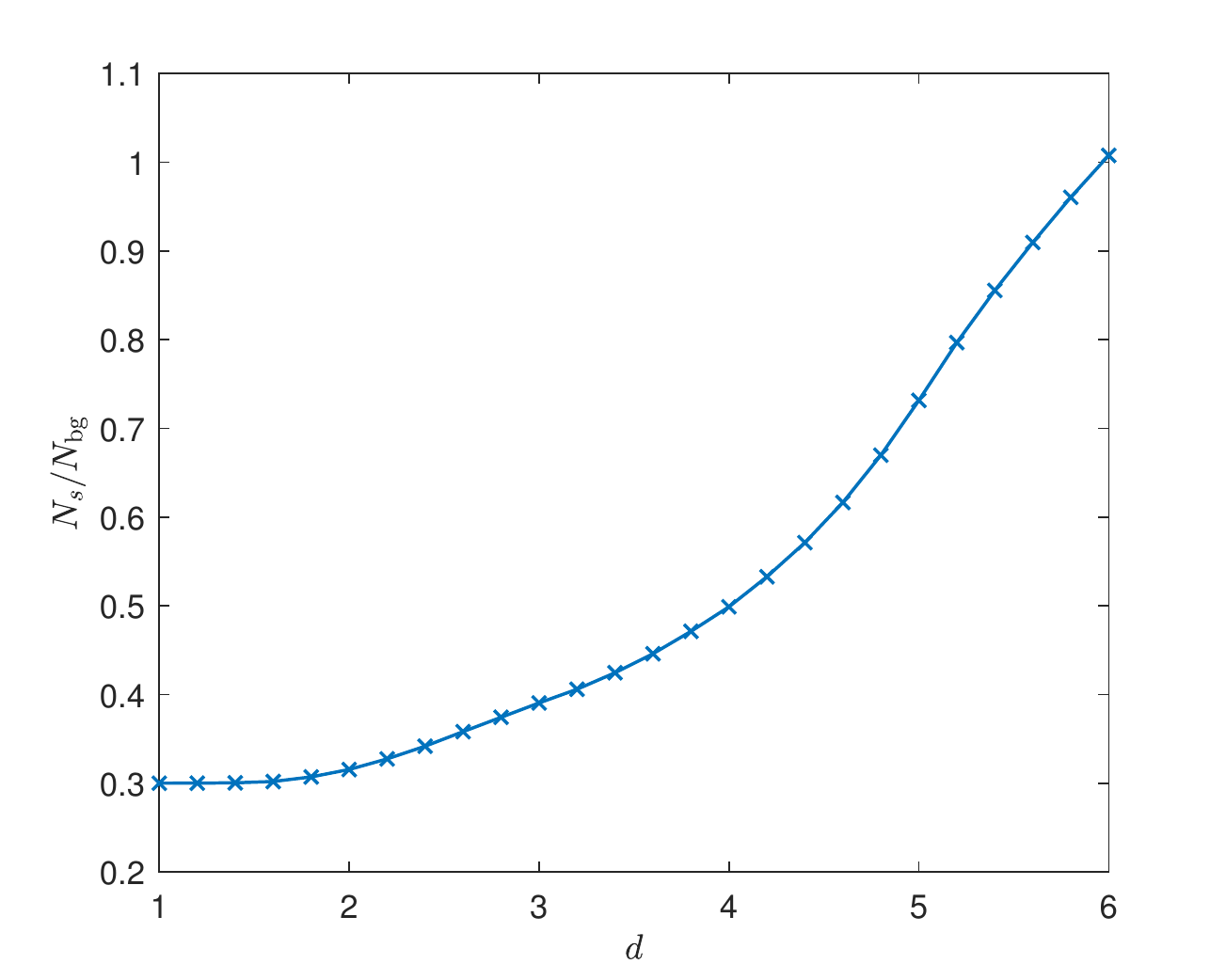}}
\caption{\label{fig:nsnbg}
$N_s/N_{\rm bg}$ in the conservative cases for $O_{T_0}$ as functions of $d$, with $f_{T_0}$ as the upper bounds of Table~\ref{table:coefficientscan}. 
The top-left panel corresponds to $\sqrt{s}=3\;{\rm TeV}$,
the top-right panel corresponds to $\sqrt{s}=10\;{\rm TeV}$,
the bottom-left panel corresponds to $\sqrt{s}=14\;{\rm TeV}$,
and the bottom-right panel corresponds to $\sqrt{s}=30\;{\rm TeV}$.}
\end{center}
\end{figure}
  
\begin{figure}[htbp]
\begin{center}
\resizebox{0.48\hsize}{!}{\includegraphics{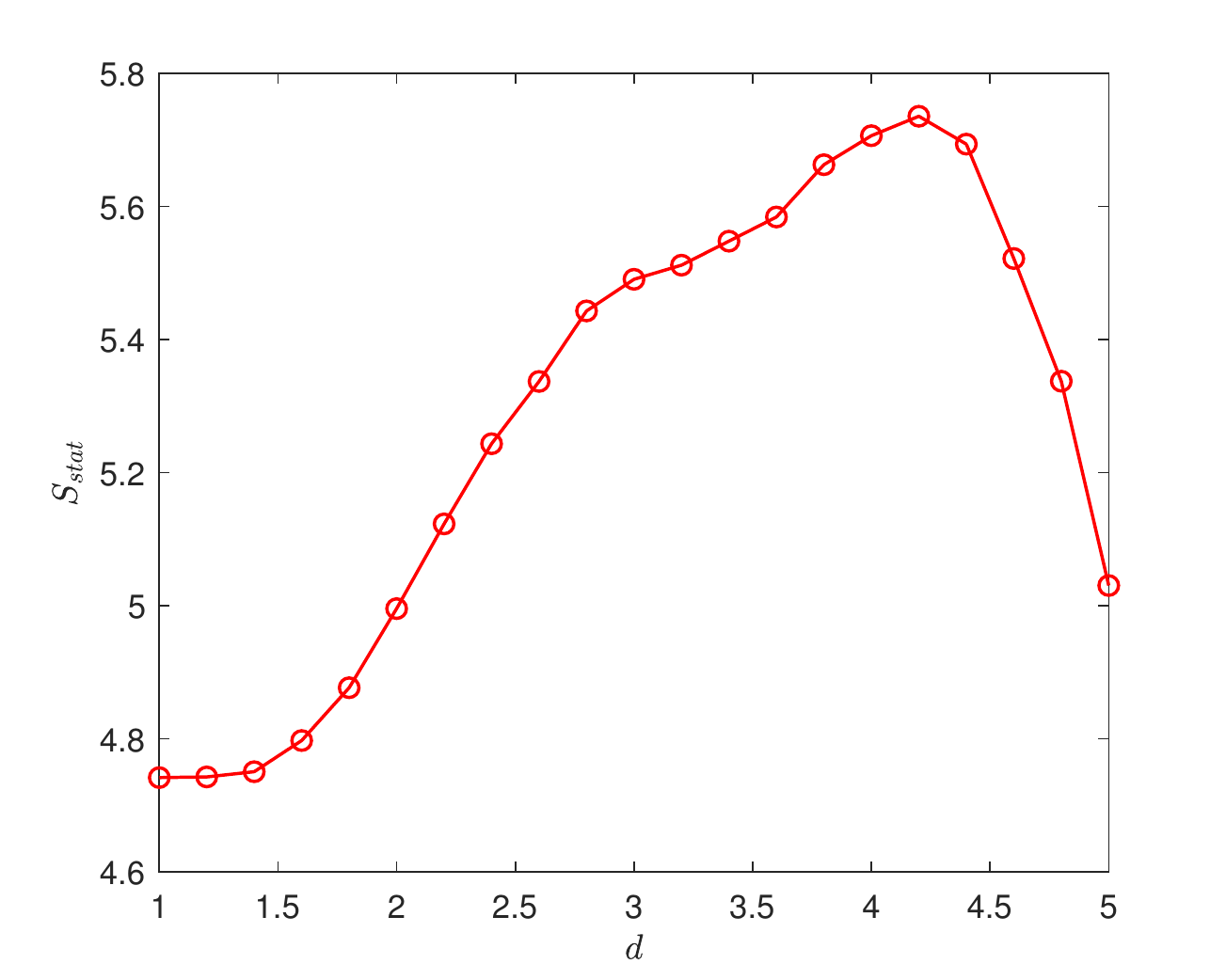}}
\resizebox{0.48\hsize}{!}{\includegraphics{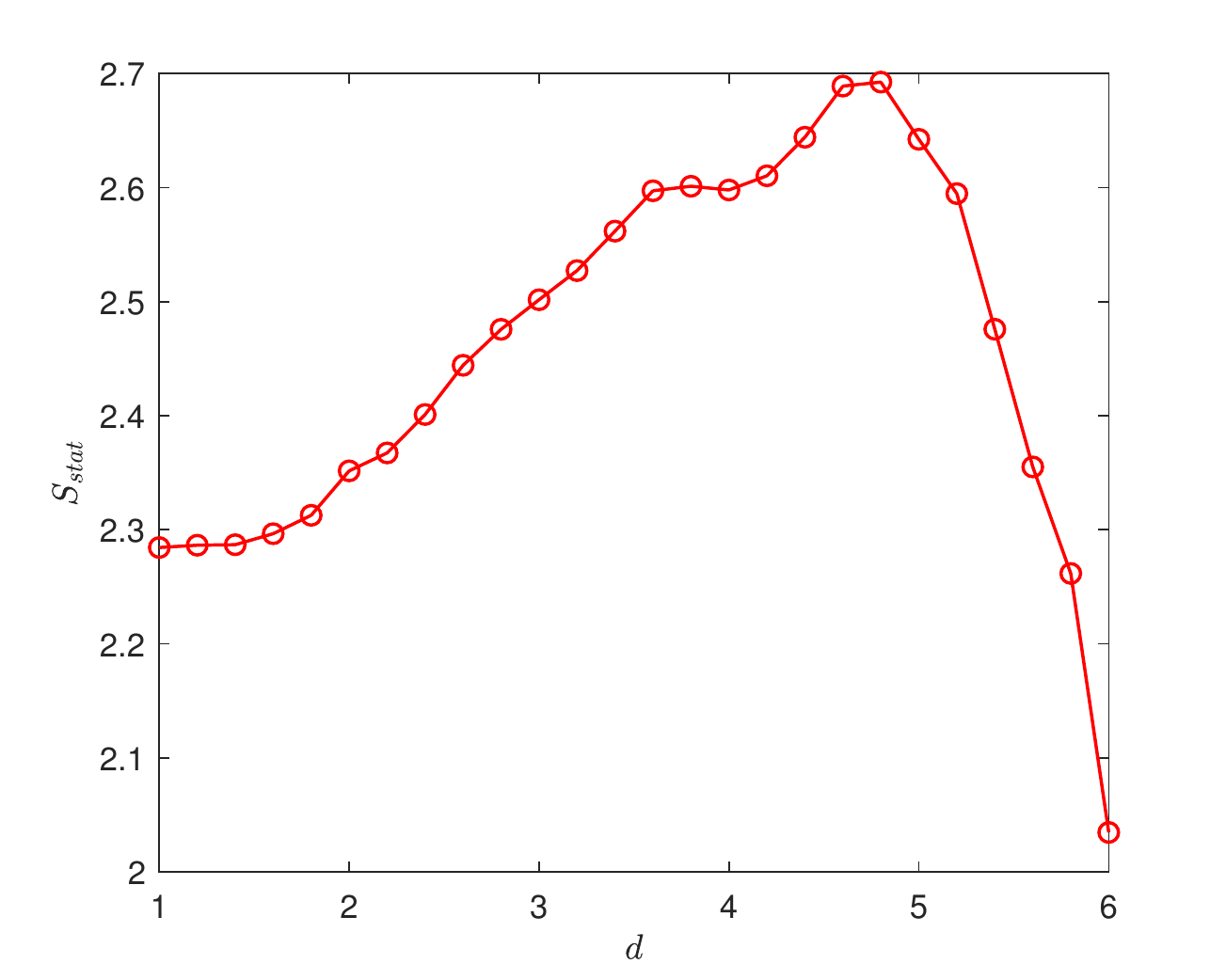}}\\
\resizebox{0.48\hsize}{!}{\includegraphics{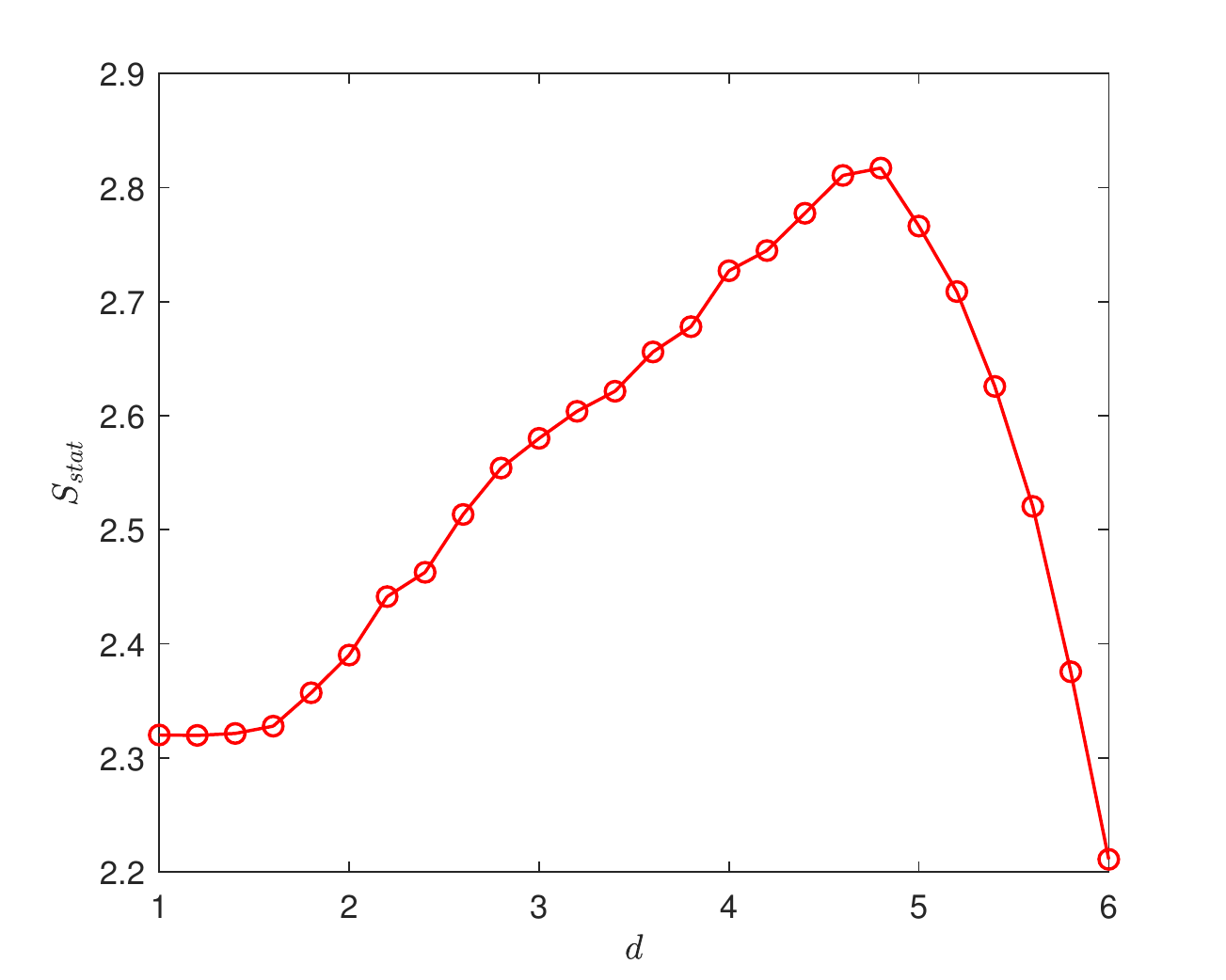}}
\resizebox{0.48\hsize}{!}{\includegraphics{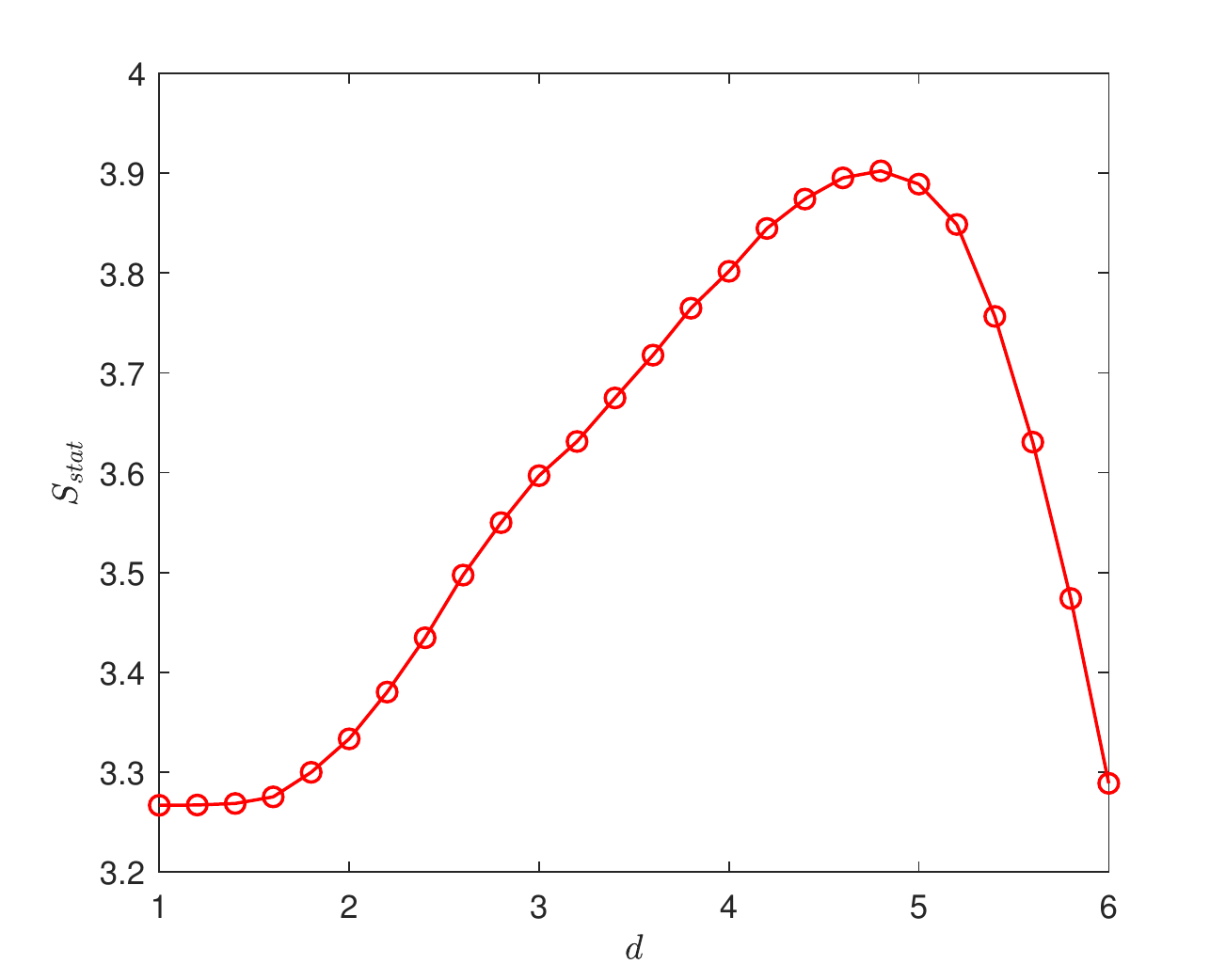}}
\caption{\label{fig:sstat}
Same as Fig.~\ref{fig:nsnbg} but for $\mathcal{S}_{\rm stat}$.}
\end{center}
\end{figure}

In order to verify the effect of different criterions, the effect of the PCAAD event selection strategy for $O_{T_0}$ at $\sqrt{s}=3$, $10$, $14$, and $30\;{\rm TeV}$ are investigated.
The cross sections after cuts corresponding to different $d$ are fitted.
The expected constraints on the coefficients are estimated with the fitted cross-sections and the statistical sensitivity defined as $\mathcal{S}_{stat}=\sqrt{2 \left[(N_{\rm bg}+N_{s}) \ln (1+N_{s}/N_{\rm bg})-N_{s}\right]}$~\cite{Cowan:2010js,pdg}, where $N_s=(\sigma-\sigma_{\rm SM})L$, $N_{\rm bg}=\sigma_{\rm SM}L$ and $L$ is the luminosity.
The luminosities correspond to the conservative case are $1$, $10$, $10$ and $10\;{\rm ab}^{-1}$ for $\sqrt{s}=3$, $10$, $14$, and $30\;{\rm TeV}$, respectively~\cite{muoncollider5}.
In the conservative case, $N_s/N_{\rm bg}$ and $\mathcal{S}_{\rm stat}$ are calculated using the fitted cross sections with the coefficients $f_{T_0}$ as the upper bounds in Table~\ref{table:coefficientscan}.
$N_s/N_{\rm bg}$ and $\mathcal{S}_{\rm stat}$ are shown in Figs.~\ref{fig:nsnbg} and \ref{fig:sstat}, respectively.
It can be seen that $S_{\rm stat}$ peeks at $d=4.2$ at $\sqrt{s}=3\;{\rm TeV}$ and $4.8$ at $\sqrt{s}=10$, $14$, and $30\;{\rm TeV}$.
As a consequence, the above $d$'s are chosen as the criterions for the PCAAD event selection strategy.

\begin{table*}[htbp]
\centering
\begin{tabular}{c|c|c|c|c|c} 
\hline
  &  &3\;{\rm TeV} &10\;{\rm TeV} &14\;{\rm TeV} &30\;{\rm TeV} \\
  &$S_{stat}$& $1\;{\rm ab}^{-1}$&$10\;{\rm ab}^{-1}$&$10\;{\rm ab}^{-1}$ &$10\;{\rm ab}^{-1}$ \\
  &  &$(10^{-2}\;{\rm TeV^{-4}})$&$(10^{-4}\;{\rm TeV^{-4}})$&$(10^{-4}\;{\rm TeV^{-4}})$&$(10^{-5}\;{\rm TeV^{-4}})$ \\
\hline
  & $2$ &$[-38.2,14.3]$&$[-31.7, 12.1]$&$[-9.01,3.93]$&$[-5.71, 3.22]$\\
$f_{T_{0}}(f_{T_{1}})/\Lambda^4$& $3$&$[-43.0,19.2]$&$[-35.8, 16.2]$&$[-10.3,5.22]$&$[-6.70, 4.21]$ \\
  & $5$ & $[-51.2,27.3]$&$[-42.7, 23.1]$&$[-12.5,7.38]$&$[-8.36,5.87]$\\
\hline
 & $2$ &$[-96.3, 23.2]$&$[-83.1,19.8]$&$[-21.0,6.55]$&$[-12.9,5.61]$ \\
$f_{T_{2}}/\Lambda^4$& $3$ &$[-105.3,32.1]$&$[-90.8,27.5]$&$[-23.4,8.94]$&$[-14.8,7.51]$ \\
  & $5$ &$[-120.7,47.5]$&$[-104.0,40.7]$&$[-27.4,13.0]$&$[-18.1,10.73]$ \\
\hline
 & $2$ &$[-9.58,2.64]$&$[-7.86, 2.24]$&$[-2.18, 0.746]$&$[-1.339, 0.626]$\\
$f_{T_{5}}(f_{T_{6}})/\Lambda^4$& $3$ &$[-10.6, 3.62]$&$[-8.69,3.07]$&$[-2.45,1.01]$&$[-1.55, 0.833]$ \\
  & $5$ &$[-12.2,5.30]$&$[-10.1,4.49]$&$[-2.89,1.46]$&$[-1.90,1.18]$ \\
\hline
&$2$ &$[-24.5,4.06]$&$[-19.0,3.38]$&$[-5.20,1.17]$&$[-3.04,1.03]$\\
$f_{T_{7}}/\Lambda^4$& $3$ &$[-26.2,5.75]$&$[-20.4,4.77]$&$[-5.66,1.63]$&$[-3.42,1.41]$ \\
  & $5$ &$[-29.2,8.75]$&$[-22.88,7.23]$&$[-6.47,2.44]$&$[-4.07,2.06]$\\
\hline
 &$2$&$[-1.55,0.424]$&$[-1.30, 0.357]$&$[-0.359, 0.120]$&$[-0.213, 0.0991]$ \\
$f_{T_{8}}/\Lambda^4$& $3$ &$[-1.71,0.582]$&$[-1.43, 0.490]$&$[-0.402, 0.163]$&$[-0.245, 0.132]$ \\
  & $5$ &$[-1.98,0.852]$&$[-1.66, 0.718]$&$[-0.475, 0.236]$& $[-0.301, 0.187]$\\
\hline
 &$2$&$[-3.99,0.648]$&$[-3.20, 0.545]$&$[-0.864, 0.189]$&$[-0.503, 0.168]$ \\
$f_{T_{9}}/\Lambda^4$& $3$ &$[-4.27,0.918]$&$[-3.43, 0.771]$&$[-0.939, 0.264]$&$[-0.565, 0.229]$\\
  & $5$ &$[-4.75,1.40]$&$[-3.83, 1.77]$&$[-1.07, 0.396]$&$[-0.671, 0.336]$\\
\hline
\end{tabular}
\caption{Expected constraints on the operator coefficients at $\sqrt{s}$ =  3\;{\rm TeV} , 10\;{\rm TeV} , 14\;{\rm TeV}, and 30\;{\rm TeV} in the conservative case.}
\label{table:constraints1}
\end{table*}

\begin{table}[htbp]
\centering
\begin{tabular}{c|c|c|c} 
\hline
  &  &14\;{\rm TeV} &30\;{\rm TeV} \\
  &$S_{stat}$& $20\;{\rm ab}^{-1}$&$90\;{\rm ab}^{-1}$ \\
  &  &$(10^{-5}\;{\rm TeV^{-4}})$&$(10^{-6}\;{\rm TeV^{-4}})$ \\
\hline
  & $2$ &$[-81.4,30.5]$&$[-39.8,14.9]$ \\
$f_{T_{0}}(f_{T_{1}})/\Lambda^4$& $3$ &$[-91.8,41.0]$&$[-44.85,20.0]$ \\
  & $5$ & $[-109.2,58.4]$&$[-53.3,28.5]$ \\
\hline
 & $2$ &$[-194.1, 49.7]$&$[-97.4,24.0]$ \\
$f_{T_{2}}/\Lambda^4$& $3$ &$[-213.0, 68.6]$&$[-106.6,33.2]$ \\
  & $5$ &$[-245.4,101.0]$&$[-122.4,49.0]$ \\
\hline
 & $2$ &$[-20.06,5.69]$&$[-9.87,2.74]$ \\
$f_{T_{5}}(f_{T_{6}})/\Lambda^4$& $3$ &$[-22.2,7.80]$&$[-10.9,3.76]$ \\
  & $5$ & $[-25.8,11.4]$&$[-12.6,5.49]$ \\
\hline
& $2$ &$[-49.0,8.69]$&$[-24.2,4.18]$ \\
$f_{T_{7}}/\Lambda^4$& $3$ &$[-52.6,12.3]$&$[-26.0,5.91]$ \\
  & $5$ &$[-58.9,18.6]$&$[-29.0,8.96]$ \\
\hline
 &$2$ &$[-3.31, 0.916]$&$[-1.57, 0.434]$ \\
$f_{T_{8}}/\Lambda^4$& $3$ &$[-3.65,1.26]$&$[-1.73, 0.595]$ \\
  & $5$ &$[-4.23,1.84]$&$[-2.00, 0.869]$ \\
\hline
 &$2$ &$[-8.16,1.41]$&$[-4.03, 0.675]$ \\
$f_{T_{9}}/\Lambda^4$& $3$ &$[-8.74,1.99]$&$[-4.31, 0.955]$ \\
  & $5$ &$[-9.77,3.02]$&$[-4.81,1.45]$ \\
\hline
\end{tabular}
\caption{Expected constraints on the operator coefficients at $\sqrt{s}$ =  14\;{\rm TeV}, and 30\;{\rm TeV} in the optimistic case.}
\label{table:constraints2}
\end{table}

In this paper, the luminosities correspond to the conservative and optimistic cases are both considered~\cite{muoncollider5}, and the expected constraints are listed in Tables~\ref{table:constraints1} and \ref{table:constraints2}, respectively. 
The energy and luminosities in this paper are the same as those used in Ref.~\cite{triphoton}, however, compared to the traditional event selection strategy used in Ref.~\cite{triphoton}, the constraints are at the same order of magnitude but generally strengthened, especially for the lower bounds.

\section{\label{sec5}Summary}

Searching for NP signals at the LHC and future colliders requires a lot of data processing.
Meanwhile, the quantum computers have great potential in processing a large amount of data. 
In this paper, we investigate the event selection strategy based on PCA algorithm, which can be accelerated by quantum computers.
We proposes a PCAAD event selection strategy based on PCA algorithm to search for NP signals.

The PCAAD is an automatic event-selection strategy that does not require a prior knowledge on the physics content of the NP. 
Since both the aQGCs and the muon colliders are of interest to HEP community, in this paper, we use the tri-photon process at muon colliders as an example.
It can be shown that PCAAD is useful and efficient in NP signal searching.
The expected upper bounds on the operator coefficients w.r.t. aQGCs are generally tighter than those obtained by a traditional event selection strategy.

\begin{acknowledgement}
This work was supported in part by the National Natural Science Foundation of China under Grants No.~12147214, the Natural Science Foundation of the Liaoning Scientific Committee No.~LJKZ0978 and the Outstanding Research Cultivation Program of Liaoning Normal University (No.~21GDL004).  
\end{acknowledgement}

\appendix
\section{\label{sec:ap1}The z-score standardization and eigenvectors used in this paper}

The means and standard deviations of $p^{\rm SM}$ are used to standardize the data-sets, which are listed in Tables~\ref{table:mean} and \ref{table:std}, respectively.
The components of the eigenvectors $\vec{\eta} _{1,2,3,4}^{\rm SM}$ are listed in Tables~\ref{table:eta1}, \ref{table:eta2}, \ref{table:eta3}, and \ref{table:eta4}, respectively.

\begin{table}[htbp]
\centering
\begin{tabular}{c|c|c|c} 
\hline
 $\sqrt{s}~{\rm (TeV)}$ & $\bar{p}^{\rm SM,1}~{\rm (GeV)}$ & $\bar{p}^{\rm SM,2}~{\rm (GeV)}$ & $\bar{p}^{\rm SM,3}~{\rm (GeV)}$  \\ 
 \hline
  $3$ & $1.452\times10^{3}$& $-0.7848$ & $1.187$  \\
\hline
  $10$ & $4.882\times10^{3}$ & $-1.853$ & $5.803$  \\ 
\hline
 $14$ & $6.846\times10^{3}$ & $1.017$ & $-1.959$  \\
\hline
$30$ & $1.471\times10^{4}$ & $17.74$ & $-2.005$  \\
\hline
$\sqrt{s}~{\rm (TeV)}$ & $\bar{p}^{\rm SM,4}~{\rm (GeV)}$ & $\bar{p}^{\rm SM,5}~{\rm (GeV)}$ & $\bar{p}^{\rm SM,6}~{\rm (GeV)}$  \\
\hline
$3$& $-4.609$&$1.313\times10^{3}$&$51.81$\\
\hline
$10$& $-24.41$&$4.511\times10^{3}$&$2.263$\\
\hline
$14$& $-29.05$&$6.354\times10^{3}$&$0.1737$\\
\hline
$30$& $-43.25$&$1.377\times10^4$&$-16.25$\\
\hline
$\sqrt{s}~{\rm (TeV)}$& $\bar{p}^{\rm SM,7}~{\rm (GeV)}$ & $\bar{p}^{\rm SM,8}~{\rm (GeV)}$& $\bar{p}^{\rm SM,9}~{\rm (GeV)}$ \\
\hline
$3$&$-0.9730$&$4.522$&$2.351\times10^{2}$\\
\hline
$10$&$-4.906$&$23.97$&$6.073\times10^{2}$\\
\hline
$14$&$2.401$&$26.96$&$8.009\times10^{2}$\\
\hline
$30$&$3.283$&$38.63$&$1.518\times10^{3}$\\
\hline
$\sqrt{s}~{\rm (TeV)}$& $\bar{p}^{\rm SM,10}~{\rm (GeV)}$ & $\bar{p}^{\rm SM,11}~{\rm (GeV)}$ & $\bar{p}^{\rm SM,12}~{\rm (GeV)}$ \\
\hline
$3$&$0.2460$&$-0.2024$&$0.1069$\\
\hline
$10$&$-0.4297$&$-0.8886$&$0.3613$\\
\hline
$14$&$-1.097$&$-0.4260$&$2.215$\\
\hline
$30$&$-1.326$&$-1.333$&$4.412$\\
\hline
\end{tabular}
\caption{The means of the SM data-sets.}
\label{table:mean}
\end{table}

\begin{table}[htbp]
\centering
\begin{tabular}{c|c|c|c} 
\hline
$\sqrt{s}$& $\epsilon ^{\rm SM,1}$ & $\epsilon ^{\rm SM,2}$ & $\epsilon ^{\rm SM,3}$  \\ 
(TeV) & (GeV) & (GeV) & (GeV) \\
 \hline
$3$&$74.78$&$5.999\times10^{2}$&$5.989\times10^{2}$\\
\hline
$10$&$2.287\times10^{2}$&$2.014\times10^{3}$&$2.010\times10^{3}$  \\ 
\hline
$14$&$3.157\times10^{2}$&$2.822\times10^{3}$&$2.823\times10^{3}$\\
\hline
$30$&$6.481\times10^{2}$&$6.052\times10^{3}$&$6.062\times10^{3}$\\
\hline
$\sqrt{s}$& $\epsilon ^{\rm SM,4}$& $\epsilon ^{\rm SM,5}$ & $\epsilon ^{\rm SM,6}$ \\
(TeV) & (GeV) & (GeV) & (GeV) \\
\hline
$3$&$1.182\times10^{3}$&$1.831\times10^{2}$&$5.696\times10^{2}$\\
\hline
$10$&$3.973\times10^{3}$&$5.856\times10^{2}$&$1.932\times10^{3}$ \\
\hline
$14$&$5.569\times10^{3}$&$8.083\times10^{2}$&$2.714\times10^{3}$\\
\hline
$30$&$1.198\times10^{4}$&$1.678\times10^{3}$&$5.840\times10^{3}$\\
\hline
$\sqrt{s}$& $\epsilon ^{\rm SM,7}$ & $\epsilon ^{\rm SM,8}$& $\epsilon ^{\rm SM,9}$\\
(TeV) & (GeV) & (GeV) & (GeV) \\
\hline
$3$&$5.686\times10^{2}$&$1.053\times10^{3}$&$2.300\times10^{2}$\\
\hline
$10$&$1.929\times10^{3}$&$3.639\times10^{3}$&$7.388\times10^{2}$\\
\hline
$14$&$2.715\times10^{3}$&$5.128\times10^{3}$&$1.023\times10^{3}$\\
\hline
$30$&$5.852\times10^{3}$&$1.114\times10^{4}$&$2.126\times10^{3}$\\
\hline
$\sqrt{s}$& $\epsilon ^{\rm SM,10}$ & $\epsilon ^{\rm SM,11}$ & $\epsilon ^{\rm SM,12}$\\
(TeV) & (GeV) & (GeV) & (GeV) \\
\hline
$3$&$1.546\times10^{2}$&$1.545\times10^{2}$&$2.458\times10^{2}$\\
\hline
$10$&$4.497\times10^{3}$&$4.487\times10^{2}$&$7.149\times10^{2}$\\
\hline
$14$&$6.106\times10^{3}$&$6.105\times10^{2}$&$9.707\times10^{2}$\\
\hline
$30$&$1.228\times10^{3}$&$1.227\times10^{3}$&$1.952\times10^{3}$\\
\hline
\end{tabular}
\caption{The standard deviations of the SM data-sets.}
\label{table:std}
\end{table}

\begin{table}[htbp]
\begin{tabular}{c|c|c|c}
\hline
$\sqrt{s}$ & $\eta^{\rm SM,1}_1$&$\eta^{\rm SM,2}_1$&$\eta^{\rm SM,3}_1$\\ 
(TeV) & & & \\
\hline
$3$&$-0.5114$&$-4.469\times10^{-3}$&$3.169\times10^{-3}$\\
\hline
$10$&$0.5220$&$-1.929\times10^{-3}$&$2.683\times10^{-3}$\\
\hline
$14$&$-0.5245$&$2.614\times10^{-3}$&$-5.238\times10^{-4}$\\ 
\hline
$30$&$-0.5273$&$3.230\times10^{-4}$&$4.392\times10^{-3}$\\
 \hline
$\sqrt{s}$&$\eta^{\rm SM,4}_1$&$\eta^{\rm SM,5}_1$&$\eta^{\rm SM,6}_1$\\
(TeV) & & & \\
\hline
$3$&$-0.5848$&$3.174\times10^{-2}$&$3.695\times10^{-3}$\\
\hline
$10$&$0.5843$&$-2.073\times10^{-2}$&$1.630\times10^{-3}$\\
\hline
$14$&$-0.5840$&$1.364\times10^{-2}$&$-1.378\times10^{-3}$\\
\hline
$30$&$-0.5839$&$-3.603\times10^{-3}$&$2.012\times10^{-4}$\\
\hline
$\sqrt{s}$&$\eta^{\rm SM,7}_1$&$\eta^{\rm SM,8}_1$&$\eta^{\rm SM,9}_1$\\
(TeV) & & & \\
\hline
$3$&$-2.569\times10^{-3}$&$-3.013\times10^{-2}$&$0.6276$\\
\hline
$10$&$-3.691\times10^{-3}$&$1.993\times10^{-2}$&$-0.6206$\\
\hline
$14$&$1.163\times10^{-3}$&$-1.360\times10^{-2}$&$0.6192$\\
\hline
$30$&$-3.748\times10^{-3}$&$3.028\times10^{-3}$&$0.6172$\\
\hline
$\sqrt{s}$&$\eta^{\rm SM,10}_1$&$\eta^{\rm SM,11}_1$&$\eta^{\rm SM,12}_1$\\
(TeV) & & & \\
\hline
$3$&$-2.569\times10^{-3}$&$-2.734\times10^{-3}$&$-2.317\times10^{-2}$\\
\hline
$10$&$-3.691\times10^{-3}$&$3.928\times10^{-3}$&$1.337\times10^{-2}$\\
\hline
$14$&$1.163\times10^{-3}$&$-2.791\times10^{-3}$&$-6.059\times10^{-3}$\\
\hline
$30$&$-3.748\times10^{-3}$&$-3.553\times10^{-3}$&$4.628\times10^{-3}$\\
\hline
\end{tabular}
\caption{Components of $\vec{\eta} _1^{\rm SM}$.}
\label{table:eta1}
\end{table}

\begin{table}[htbp]
\begin{tabular}{c|c|c|c}
\hline
$\sqrt{s}$ & $\eta^{\rm SM,1}_2$&$\eta^{\rm SM,2}_2$&$\eta^{\rm SM,3}_2$\\ 
(TeV) & & & \\
\hline
$3$&$-2.533\times10^{-2}$&$-4.052\times10^{-3}$&$1.522\times10^{-3}$\\
\hline
$10$&$1.670\times10^{-2}$&$7.755\times10^{-3}$&$5.535\times10^{-3}$\\
\hline
$14$&$-9.759\times10^{-3}$&$-4.380\times10^{-3}$&$8.598\times10^{-4}$\\ 
\hline
$30$&$3.493\times10^{-3}$&$-4.147\times10^{-3}$&$-4.742\times^10{-3}$\\
 \hline
$\sqrt{s}$&$\eta^{\rm SM,4}_2$&$\eta^{\rm SM,5}_2$&$\eta^{\rm SM,6}_2$\\
(TeV) & & & \\
\hline
$3$&$-0.6408$&$-2.900\times10^{-2}$&$4.151\times10^{-3}$\\
\hline
$10$&$0.6502$&$1.849\times10^{-2}$&$-7.562\times10^{-3}$\\
\hline
$14$&$-0.6528$&$-1.222\times10^{-2}$&$4.239\times10^{-3}$\\
\hline
$30$&$-0.6571$&$3.336\times10^{-3}$&$3.648\times10^{-3}$\\
\hline
$\sqrt{s}$&$\eta^{\rm SM,7}_2$&$\eta^{\rm SM,8}_2$&$\eta^{\rm SM,9}_2$\\
(TeV) & & & \\
\hline
$3$&$-1.472\times10^{-3}$&$0.6086$&$3.107\times10^{-2}$\\
\hline
$10$&$-5.162\times10^{-3}$&$-0.6216$&$-1.962\times10^{-2}$\\
\hline
$14$&$-8.473\times10^{-4}$&$0.6251$&$1.255\times10^{-2}$\\
\hline
$30$&$4.691\times10^{-3}$&$0.6312$&$-3.801\times10^{-3}$\\
\hline
$\sqrt{s}$&$\eta^{\rm SM,10}_2$&$\eta^{\rm SM,11}_2$&$\eta^{\rm SM,12}_2$\\
(TeV) & & & \\
\hline
$3$&$3.748\times10^{-4}$&$-4.076\times10^{-4}$&$0.4654$\\
\hline
$10$&$-1.933\times10^{-3}$&$-2.469\times10^{-3}$&$-0.4356$\\
\hline
$14$&$1.375\times10^{-3}$&$-1.196\times10^{-4}$&$0.4274$\\
\hline
$30$&$2.920\times10^{-3}$&$1.048\times10^{-3}$&$0.4120$\\
\hline
\end{tabular}
\caption{Components of $\vec{\eta} _2^{\rm SM}$.}
\label{table:eta2}
\end{table}

\begin{table}[htbp]
\begin{tabular}{c|c|c|c}
\hline
$\sqrt{s}$ & $\eta^{\rm SM,1}_3$&$\eta^{\rm SM,2}_3$&$\eta^{\rm SM,3}_3$\\ 
(TeV) & & & \\
\hline
$3$&$-4.063\times10^{-4}$&$-0.3922$&$-0.5816$\\
\hline
$10$&$5.089\times10^{-4}$&$0.5243$&$0.4680$\\
\hline
$14$&$-1.523\times10^{-3}$&$-0.5086$&$0.4855$\\ 
\hline
$30$&$-2.027\times10^{-3}$&$-0.6127$&$-0.3457$\\
 \hline
$\sqrt{s}$&$\eta^{\rm SM,4}_3$&$\eta^{\rm SM,5}_3$&$\eta^{\rm SM,6}_3$\\
(TeV) & & & \\
\hline
$3$&$8.379\times10^{-4}$&$1.516\times10^{-4}$&$0.3743$\\
\hline
$10$&$-9.054\times10^{-3}$&$-9.279\times10^{-4}$&$-0.5066$\\
\hline
$14$&$3.844\times10^{-3}$&$-1.924\times10^{-3}$&$0.4930$\\
\hline
$30$&$5.603\times10^{-3}$&$-2.072\times10^{-3}$&$0.5959$\\
\hline
$\sqrt{s}$&$\eta^{\rm SM,7}_3$&$\eta^{\rm SM,8}_3$&$\eta^{\rm SM,9}_3$\\
(TeV) & & & \\
\hline
$3$&$0.5556$&$-7.087\times10^{-4}$&$-7.999\times10^{-6}$\\
\hline
$10$&$-0.4530$&$8.961\times10^{-3}$&$7.055\times10^{-4}$\\
\hline
$14$&$-0.4704$&$-3.942\times10^{-3}$&$2.016\times10^{-3}$\\
\hline
$30$&$0.3359$&$-5.385\times10^{-3}$&$2.254\times10^{-3}$\\
\hline
$\sqrt{s}$&$\eta^{\rm SM,10}_3$&$\eta^{\rm SM,11}_3$&$\eta^{\rm SM,12}_3$\\
(TeV) & & & \\
\hline
$3$&$0.1370$&$0.2008$&$-9.601\times10^{-4}$\\
\hline
$10$&$-0.1585$&$-0.1379$&$4.574\times10^{-3}$\\
\hline
$14$&$0.1466$&$-0.1405$&$-1.035\times10^{-3}$\\
\hline
$30$&$0.1672$&$9.542\times10^{-2}$&$-3.435\times10^{-3}$\\
\hline
\end{tabular}
\caption{Components of $\vec{\eta} _3^{\rm SM}$.}
\label{table:eta3}
\end{table}

\begin{table}[htbp]
\begin{tabular}{c|c|c|c}
\hline
$\sqrt{s}$ &$\eta^{\rm SM,1}_4$&$\eta^{\rm SM,2}_4$&$\eta^{\rm SM,3}_4$\\ 
(TeV) & & & \\
\hline
$3$&$4.182\times10^{-3}$&$-0.5817$&$0.3924$\\
\hline
$10$&$2.540\times10^{-3}$&$0.4679$&$-0.5244$\\
\hline
$14$&$2.148\times10^{-3}$&$0.4855$&$0.5087$\\ 
\hline
$30$&$-2.128\times10^{-3}$&$0.3457$&$-0.6129$\\
 \hline
$\sqrt{s}$&$\eta^{\rm SM,4}_4$&$\eta^{\rm SM,5}_4$&$\eta^{\rm SM,6}_4$\\
(TeV) & & & \\
\hline
$3$&$3.880\times10^{-3}$&$4.826\times10^{-3}$&$0.5568$\\
\hline
$10$&$-6.008\times10^{-4}$&$2.530\times10^{-3}$&$-0.4525$\\
\hline
$14$&$-2.485\times10^{-3}$&$7.306\times10^{-4}$&$-0.4701$\\
\hline
$30$&$2.178\times10^{-3}$&$-3.460\times10^{-3}$&$-0.3359$\\
\hline
$\sqrt{s}$&$\eta^{\rm SM,7}_4$&$\eta^{\rm SM,8}_4$&$\eta^{\rm SM,9}_4$\\
(TeV) & & & \\
\hline
$3$&$-0.3758$&$4.033\times10^{-3}$&$-5.192\times10^{-3}$\\
\hline
$10$&$0.5066$&$1.934\times10^{-4}$&$-2.696\times10^{-3}$\\
\hline
$14$&$-0.4930$&$2.562\times10^{-3}$&$-1.125\times10^{-3}$\\
\hline
$30$&$0.5965$&$-2.309\times10^{-3}$&$3.399\times10^{-3}$\\
\hline
$\sqrt{s}$&$\eta^{\rm SM,10}_4$&$\eta^{\rm SM,11}_4$&$\eta^{\rm SM,12}_4$\\
(TeV) & & & \\
\hline
$3$&$0.1968$&$-0.1322$&$-1.287\times10^{-3}$\\
\hline
$10$&$-0.1398$&$0.1583$&$2.397\times10^{-3}$\\
\hline
$14$&$-0.1417$&$-0.1465$&$6.114\times10^{-4}$\\
\hline
$30$&$-9.577\times10^{-2}$&$0.1645$&$-4.454\times10^{-5}$\\
\hline
\end{tabular}
\caption{Components of $\vec{\eta} _4^{\rm SM}$.}
\label{table:eta4}
\end{table}

\bibliography{pca}
\bibliographystyle{elsarticle-num}

\end{document}